\documentclass[12pt,preprint]{aastex}
\usepackage{epsf}
\usepackage{natbib}
\usepackage{float}
\usepackage{appendix}
\usepackage{amsmath}
\usepackage{pdflscape}
\usepackage{gensymb}
\usepackage{afterpage}

\tabletypesize{\scriptsize}

\begin{document}
\shorttitle{The SPF of Solar System Debris}
\shortauthors{Hedman \& Stark}

\title{Saturn's G and D rings provide nearly complete \\ 
measured scattering/phase functions of nearby debris disks}

\author{Matthew M. Hedman\altaffilmark{1}, Christopher C. Stark\altaffilmark{2}}

\altaffiltext{1}{Physics Department, University of Idaho, 875 Perimeter Drive MS 0903, Moscow, ID 83844-0903, USA; mhedman@uidaho.edu}
\altaffiltext{2}{Space Telescope Science Institute, 3700 San Martin Dr, Baltimore, MD 21218; cstark@stsci.edu}

\begin{abstract}
The appearance of debris disks around distant stars depends upon the scattering/phase function (SPF) of the material in the disk. However, characterizing the SPFs of these extrasolar debris disks is challenging because only a limited range of scattering angles are visible to Earth-based observers. By contrast, Saturn's tenuous rings can be observed over a much broader range of geometries, so their SPFs can be much better constrained. Since these rings are composed of small particles released from the surfaces of larger bodies, they are reasonable analogs to debris disks and so their SPFs can provide insights into the plausible scattering properties of debris disks. This work examines two of Saturn's dusty rings: the G ring (at 167,500 km from Saturn's center) and the D68 ringlet (at 67,600 km).  Using data from the cameras onboard the Cassini spacecraft, we are able to estimate the rings' brightnesses at scattering angles ranging from 170$^\circ$ to 0.5$^\circ$. We find that both of the rings exhibit extremely strong forward-scattering peaks, but for scattering angles above 60$^\circ$ their brightnesses are nearly constant. These SPFs can be well approximated by a linear combination of three Henyey-Greenstein functions, and are roughly consistent with the SPFs of irregular particles from laboratory measurements. Comparing these data to Fraunhofer and Mie models highlights several challenges involved in extracting information about particle compositions and size distributions from SPFs alone. The SPFs of these rings also indicate that the degree of forward scattering in debris disks may be greatly underestimated.
 \end{abstract}


\section{Introduction}
\label{introduction}

The material in extrasolar debris disks consists of fine particles lofted from the surfaces of  planetesimals by collisions and other processes. This debris can exhibit a wide variety of structures, including rings \citep[e.g.,][]{kalas2005,schneider2006}, warps \citep[e.g.,][]{golimowski2006, krist2005}, and brightness asymmetries \citep[e.g.,][]{hines2007,kalas2007}.  One of the most common and pronounced asymmetries in moderately-inclined debris disks is a brightness inequality along the projected minor axis \citep[e.g.,][]{schneider2014}.  
This asymmetry is likely the result of anisotropic ``forward" scattering of starlight by dust grains, with the near side of the disk appearing brighter than the far side. 

Brightness variations due to the anisotropic scattering properties of the circumstellar debris can be quantified by the material's scattering/phase function (SPF). This function specifies the relative brightness of the material as a function of either the phase angle $\alpha$ (i.e. the angle between the rays followed by the incident and scattered starlight) or its supplement, the scattering angle $\theta$. If the SPF of the disk material is sufficiently well known, then the brightness variations due to the changing lighting geometry could be removed, revealing the density variations that could be generated by unseen planets \citep{stark2014}.  The SPF can also be used to constrain the apparent albedo, and thus the mass, of a disk.  Finally, if debris disks are strongly forward-scattering as recent observations suggest \citep{stark2014,perrin2014}, the forward-scattered starlight from regions more than 5 AU from the star may create additional pseudo-zodiacal light in edge-on systems with which future exoEarth-imaging missions must contend \citep{stark2015}.  The SPF of debris dust therefore plays a critical role in determining the dynamical state of a planetary system, constraining its composition, and designing future missions.

Unfortunately, the SPFs of the material in exoplanetary systems are poorly constrained. To determine the SPF, one must observe the disk over a large range of scattering angles, especially those near the forward-scattering peak.  For debris disks not observed edge-on,  the range of scattering angles that nature provides are limited to a range of values around $90^\circ$ that depends upon the inclination of the system. Currently-available observations of debris disks reveal that their SPFs vary slowly at scattering angles around $90^{\degree}$, with typical Henyey-Greenstein fits adopting values of $0.0 < g \lesssim 0.3$ \citep[e.g.][] {kalas2005,schneider2006,debes2008,thalmann2011}.  These results are somewhat surprising, given that classic Mie theory predicts highly forward-scattering dust, with the first moment of the SPF, $\langle\cos{\theta}\rangle \sim 0.9$. While there are SPFs that exhibit both a strong forward-scattering peak and a flat phase function near $\theta\sim90\degree$, these findings still highlight the difficulties involved in extrapolating the measured SPF of exoplanetary disks beyond the narrow observed range of phase angles. Larger ranges of phase angles are potentially observable in edge-on disks, but in these situations there are still limitations imposed by the inner working angle of the observations. Furthermore, in these systems the SPF becomes degenerate with the radial variations in the dust density and size distribution. Estimates of our local zodiacal cloud's scattering phase function suffer from these same degeneracies \citep[e.g.,][]{hong1985}.

In lieu of being able to measure the complete SPF of debris disks directly, it is useful to consider dusty tenuous rings that surround the giant planets. Unlike the famous dense rings of Saturn, which are composed mostly of pebble-to-boulder-sized chunks of ice \citep[see][and references therein]{cuzzi2009}, these much fainter and more tenuous rings appear to contain much smaller particles. Indeed, these tenuous rings appear much brighter when viewed at high phase angles, which indicates that most of the visible particles are less than 100 microns across. Such small particles will be destroyed or ejected from the planetary system on timescales less than a few thousand years  \citep{burns2001, horanyi2009}, so they need to be continuously supplied to the ring from larger objects. Indeed, many tenuous rings contain small moons that are likely the largest of those source bodies, and visible dusty ring particles probably consist of material released from the surfaces of these objects by collisions with meteoroids, much like how the material in debris disks is thought to arise from the collisions among planetesimals in orbit around the star. Furthermore, these dusty rings have been observed over a broad range of viewing geometries by spacecraft, yielding a much more complete picture of their SPFs. Analyses of Voyager and ground-based data yielded sparse phase curves of both Saturn's E and G rings \citep{showalter1991, showalter1993, throop1998}, but even these were enough to clearly show the forward-scattering peak. More recently, the combined data from multiple missions have provided well-sampled phase curves of both Jupiter's main ring \citep{throop2004} and Saturn's F ring \citep{french2012}, both of which show a clear forward-scattering peak and a relatively flat SPF for phase angles less than 120$^\circ$. However, both rings show significant longitudinal and/or temporal variations in their brightness, which complicates efforts to interpret the details of these SPFs.

\begin{figure}
\resizebox{6.5in}{!}{\includegraphics{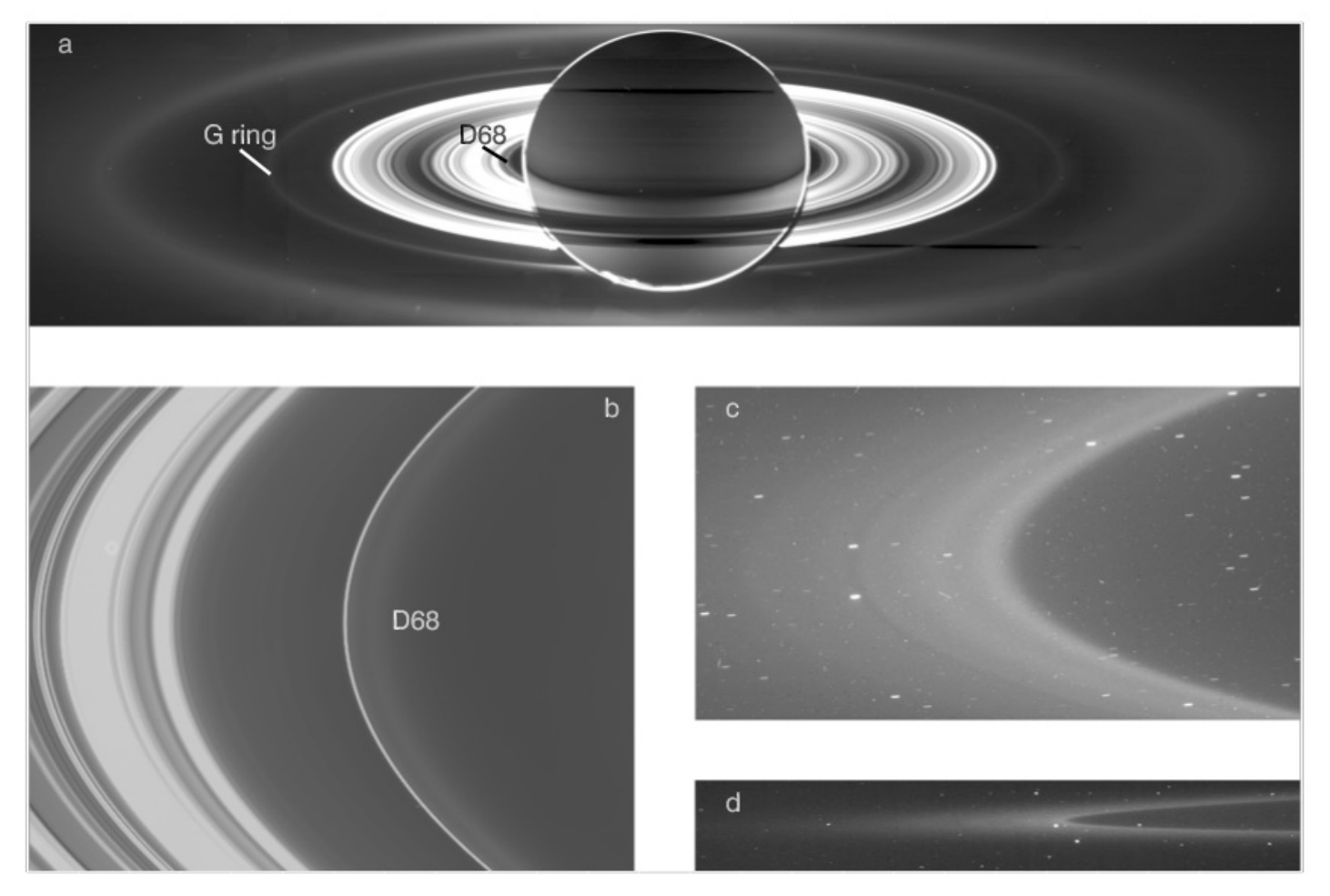}}
\caption{The general appearance and location of the G ring and D68 within Saturn's rings. All images have been individually stretched and have been rotated so that Saturn's north pole would point upwards. (a) A mosaic of images obtained  by the Cassini spacecraft in September 2006 during a time when the spacecraft flew through Saturn's shadow. At these high phase angles, both the G ring outside the main rings and D68 in the inner D ring can be clearly seen (note this image uses a gamma stretch so that a broad range of brightnesses is visible). (b) A close-up of the D68 ringlet from Cassini image N1537019704, obtained during the same time as the above maosaic. Note the narrow ringlet is unresolved in this image. (c) and (d) The G ring at two different opening angles from Cassini images  N1643025601 (phase angle 130.5$^\circ$, ring opening angle 3.9$^\circ$) and N1634622552 (phase angle 100.6$^\circ$, ring opening angle 0.5$^\circ$). Note the ring's  relatively sharp inner edge and more diffuse outer boundary.}
\label{dgcontext}
\end{figure}

In this paper we will use data from the cameras onboard Cassini spacecraft to constrain the SPF of two of Saturn's tenuous dusty rings over a range of scattering angles from 170$^\circ$ to 0.5$^\circ$ (i.e. phase angles from  10$^\circ$ to $179.5^\circ$), thereby probing further up the forward-scattering peak than ever before. Specifically, we will examine here the inner edge of the G ring and a narrow ringlet called D68 in the D ring. The G ring is located at approximately 167,500 km from Saturn's center, outside Saturn's main ring system  (see Figure~\ref{dgcontext}). This ring probably consists of micrometeorite-ejected debris from larger objects (including a small moon called Aegaeon) confined near the inner edge of the ring by a co-rotation resonance with Mimas \citep{showalter1993, lissauer2000, hedman2007, hedman2010}. D68, on the other hand, is found only 67,600 km from Saturn's center (see Figure~\ref{dgcontext}) and is the innermost discrete ringlet in the ring system  \citep{showalter1996, hedman2007d, hedman2014}. Unlike the G ring, there is no obvious local source of dust in this region. However, D68 is embedded in a broad sheet of dust extending interior to Saturn's main rings, and so it probably represents material that has become trapped as it drifted inwards towards the planet. While Saturn possesses many other dusty rings, we have chosen to focus of these two because they are comparatively narrow, which means the relevant signals can be more easily extracted from instrumental and astronomical backgrounds. Furthermore, while both these rings exhibit some brightness variations, neither one is as obviously clumpy and time-variable as the F ring \citep{french2012}, or as strongly asymmetric as the E ring \citep{hedman2012}. These two rings therefore promise to provide reasonably coherent phase functions that could be useful for modeling exoplanetary debris disks. 

In Section \ref{observations} we describe the Cassini observations and reduction techniques, as well as our methods for measuring the SPF.  In Section \ref{results} we present the measured scattering phase functions for D68 and the G ring, while in Section~\ref{fits} we attempt to fit these SPFs to Henyey-Greenstein functions, laboratory-measured phase functions, and models based on Faunhofer diffraction and Mie theory.  We discuss some implications of the measured SPFs for exoplanetary disks in Section \ref{discussion} and summarize our findings in Section \ref{conclusions}.

\section{Observations \& Data Reduction}
\label{observations}

\subsection{Data sources}

The raw data for this investigation consists of images obtained by the Imaging Science Subsystem (ISS) onboard the Cassini Spacecraft. We performed a comprehensive search for images of both the G ring and D68, and then selected images with sufficient signal to noise to detect these ring features. This search included images obtained by both the Narrow Angle Camera (NAC) and the Wide Angle Camera (WAC) components of the Imaging Science Subsystem \citep{porco2004, west2010}. Both of these cameras have multiple filters, but for the purposes of this analysis we only considered images obtained through the clear filters or the RED filter on the WAC. Note that the effective wavelength of the RED filter (647 nm) is close to the effective wavelengths of the NAC and WAC clear filters (651 nm and 634 nm, respectively, Porco {\it et al.} 2004). Thus the measured brightness through these different filters should be insensitive to the rings' spectral slopes. Each image was calibrated using the standard CISSCAL calibration routines \citep{porco2004, west2010}, which remove instrumental backgrounds, flat-field the image and convert the raw data numbers into $I/F$, a measure of surface reflectance that is unity for a Lambertian surface viewed and illuminated at normal incidence (see {\tt http://pds-rings.seti.org/cassini/iss/calibration.html}). Each image was also geometrically navigated using the appropriate SPICE kernels \citep{acton1996}, and the detailed geometry of the image was refined based on the locations of stars and/or sharp ring edges in the field of view. 

\subsection{From images to radial profiles}

The data from these images were reduced to profiles of the ring's observed brightness as a function of ring radius (distance from the planet's spin axis). For images obtained at scattering angles greater than 15$^\circ$ (i.e. phase angles less than 165$^\circ$) and ring opening angles greater than 0.5$^\circ$, these profiles were computed by simply averaging the brightness in the image over a range of longitudes, avoiding any part of the ring that was in shadow or had low radial resolution due to projection effects. During this process, we also compute the scattering angle at the ring $\theta$ and the opening angle of the ring to the spacecraft $B$.  The ring opening angle is relevant to this analysis because the total path length through the ring along the line of sight is proportional to $1/\sin|B|$. Hence, for any tenuous ring the total amount of material along the line of sight and the apparent brightness of the ring are proportional to $1/\sin|B|$. We therefore multiply the observed ring brightness by $\sin|B|$ to convert the observed $I/F$ into a ``normal $I/F$" (also denoted $\mu I/F$). For low optical depth rings like the G and D rings, multiplying through by this factor removes the dependance on $B$ (indeed, the normal $I/F$ can be regarded as the $I/F$ that would be measured if the ring was viewed face-on, with $B=90^\circ$) and so this quantity should be independent of viewing geometry for a given scattering angle.

A small number of G-ring images were obtained when the spacecraft was less  than 0.5$^\circ$ from the ringplane (see for example Figure~\ref{dgcontext}d). In these nearly edge-on images, the finite vertical extent of the ring can influence the detailed shape of a brightness profile derived using the above methods, and these changes can interfere with efforts to quantify the ring's brightness. Hence we use a different set of procedures for these images. First, we sum the $I/F$ values along pixel columns orthogonal to the radial direction to produce a profile of the vertically integrated $I/F$ versus radius. In such a profile, the signal at a given radius $r_0$  includes contributions from material at all radii $r>r_0$. However, these projection effects can be removed using an onion-peeling algorithm, which iteratively estimates the signal at each radius and removes that material's contribution from the rest of the data \citep{showalter1985, showalter1987, depater2004, hedman2012}. When properly normalized, this algorithm yields a profile of the ring's normal $I/F$ that can be directly compared with the data obtained at higher elevation angles. However, the repeated differencing of the vertically-integrated $I/F$ profile also amplifies small noise fluctuations in the final brightness data, so we only use this method on the lowest-elevation images where other methods of generating the brightness profile clearly distort the rings' shape. 

We also used different reduction procedures on images where the rings were viewed at scattering angles less than 15$^\circ$ (phase angles greater than 165$^\circ$). In these cases, the SPF of the relevant rings is very steep and so the brightness of the ring could vary significantly over the range of scattering angles visible in a single image. Hence, instead of reducing each image to a single profile, we extracted multiple brightness profiles from each image. This was done in slightly different ways for the two ring features.

For the narrow D68 ringlet, the low-scattering-angle data consist primarily of individual images that captured an entire ring ansa (like that shown in Figure~\ref{dgcontext}b). Thus multiple profiles were obtained from each image by simply averaging together different ranges of ring longitudes within each image. 

\begin{figure}
\resizebox{6.5in}{!}{\includegraphics{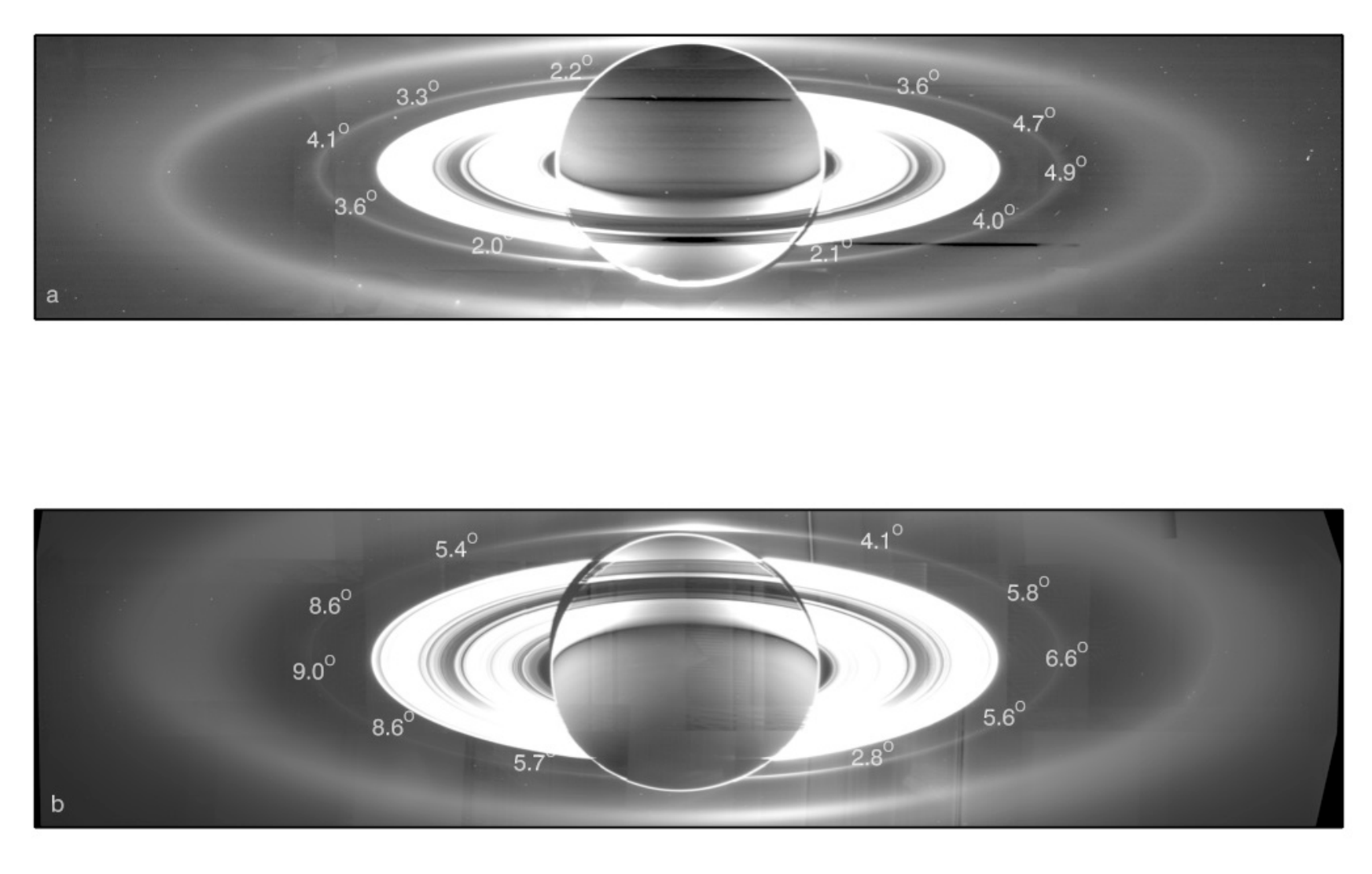}}
\caption{Two of the mosaics obtained when Cassini flew through Saturn's shadow: (a) is the Rev 028 HIPHWAC sequence from 2006, while (b) is the Rev 195 HIPHASEC sequence from 2013. The scattering angle (180$^\circ$-phase angle) of various locations in the G ring are given by the numbers. Because the Sun is behind Saturn in these mosaics, there are up to four different locations in the ring that can have the same scattering angle. We consider the four quadrants of the ring separately when computing profiles. Note that profiles are generated by combining data from the individual  images, and not from mosaics like these (which are created only for illustration purposes).}
\label{gphaselab}
\end{figure}

For the more diffuse G ring, individual images were less likely to capture a range of radii and longitudes needed to yield suitable ring profiles. Thus we instead considered four observation sequences that were particularly informative about the rings' scattering/phase function. Three of these sequences (Rev 028\footnote{``Rev" is a designation for one of Cassini's orbit around Saturn.} HIPHWAC, Rev 173 HIPHWAC and Rev 195 HIPHASEC) were mosaics of many images that together captured nearly the entire ring, while the fourth observation (Rev 028 HIPHASE002) was a movie where the camera observed the brightness of one part of the G ring change as the Sun moved relative to the observation point. Since the relevant data could be spread across multiple images, for these observations it made more sense to compute profiles by averaging together data from several images covering a restricted range of phase angles rather than just consider a limited range of longitudes in each image. While this procedure was straightforward for the movie sequence, for the three large-scale mosaics there was an added complication; between two and four disjoint portions of the ring could have the same phase angle (see Figure~\ref{gphaselab}). We therefore separated the relevant datasets into four ``quadrants'' prior to computing the averaged profiles.

\subsection{From Profiles to Brightness Estimates}

We used automated procedures the extract estimates of the relevant ring's brightness from each of the profiles derived above. Given the differences in the shapes and environments of D68 and the G ring, different algorithms were employed to quantify the brightness of the two ring structures.

\begin{figure}
\centerline{\resizebox{4in}{!}{\includegraphics{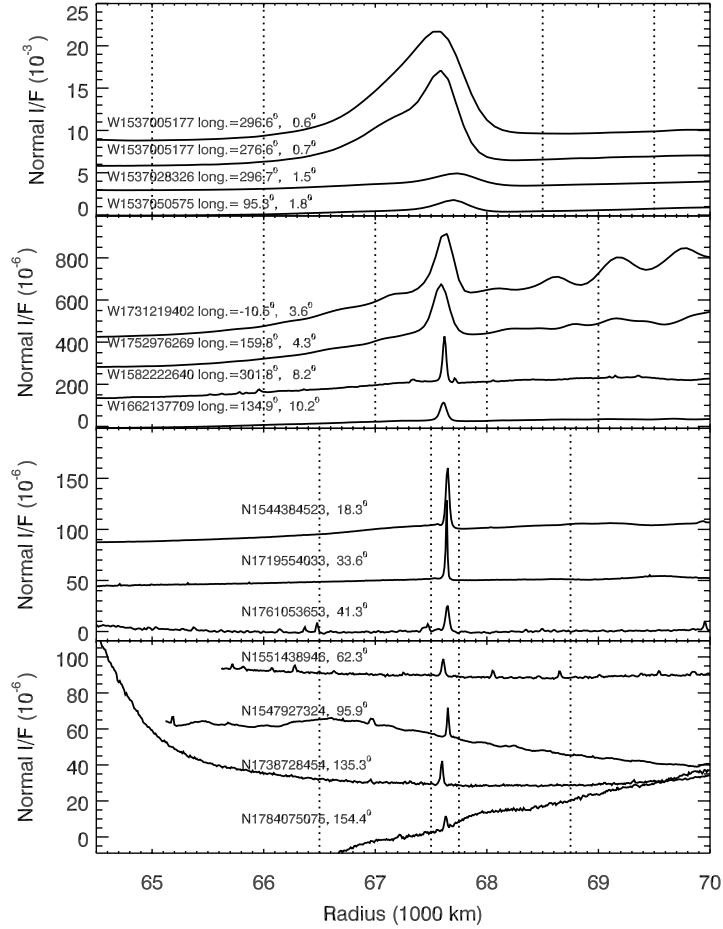}}}
\caption{Representative profiles of D68 derived from various Cassini images. Each profile is labeled with the Cassini image number and the scattering angle of the ring in that image (for the profiles with scattering angles less than 15$^\circ$, the average longitude of the data used for each profile).  Profiles are vertically offset for clarity, and different panels use different vertical scales to cope with the large variations in the ringlet's brightness with scattering angle.  In all these profiles the D68 is the peak around 67,600 km, and the width of this peak is determined primarily by the resolution of the relevant image. The vertical dotted lines in each panel indicate the range of radii flanking the ringlet that were used to determine the background behind the ringlet and the central region where the data were fit or integrated to determine the ringlet's brightness.}
\label{d68prof}
\end{figure}

D68 is a relatively isolated ringlet embedded in a rather homogeneous sheet of material (see Figure~\ref{d68prof}), and the ringlet itself is unresolved in almost all Cassini images \citep{hedman2007d, hedman2014}. Quantities like the ringlet's peak brightness therefore depend on the image resolution and are not ideal for this investigation. Hence we instead quantify the brightness of this feature in terms of its normal equivalent width (NEW), which is the ringlet's  radially-integrated normal $I/F$ above any smoothly varying background level. This parameter is independent of image resolution and elevation angle for any ring with sufficiently low optical depth. In practice, we use two different methods to compute the normal equivalent width of D68 from each profile:
\begin{itemize}
\item 
{\bf Int Method:} The brightness of a selected part of the brightness profile is directly integrated after removing a quadratic background fit to two 1000-km-wide zones on either side of the selected region  (see Figure~\ref{d68prof}).  For scattering angles greater than 15$^\circ$, the selected region ran from 67,500 km to 67,750 km, and so the background was based on a fit to the zones between 66,500-67,500 km and 67,750-68,750 km. Due to differences in the image resolution and the appearance of the ringlet, different ranges were used for profiles obtained at low scattering angles. For scattering angles between 3$^\circ$ and 15$^\circ$, the integral was compute for the region between 67,000-68,000 km, and for scattering angles below 3$^\circ$, the region was 66,000-68,500 km. 
\item 
{\bf Fit Method:} The profile of D68 is fit to a Lorentzian peak plus linear background. The integrated brightness under the peak is then determined from the product of the peak amplitude and the peak width.  This fit was performed on the data within the same radial range that was integrated over with the Int Method. 
\end{itemize}
As discussed in more detail below, these two methods generally yield similar estimates of the ringlet's normal equivalent width.

\begin{figure}
\centerline{\resizebox{4in}{!}{\includegraphics{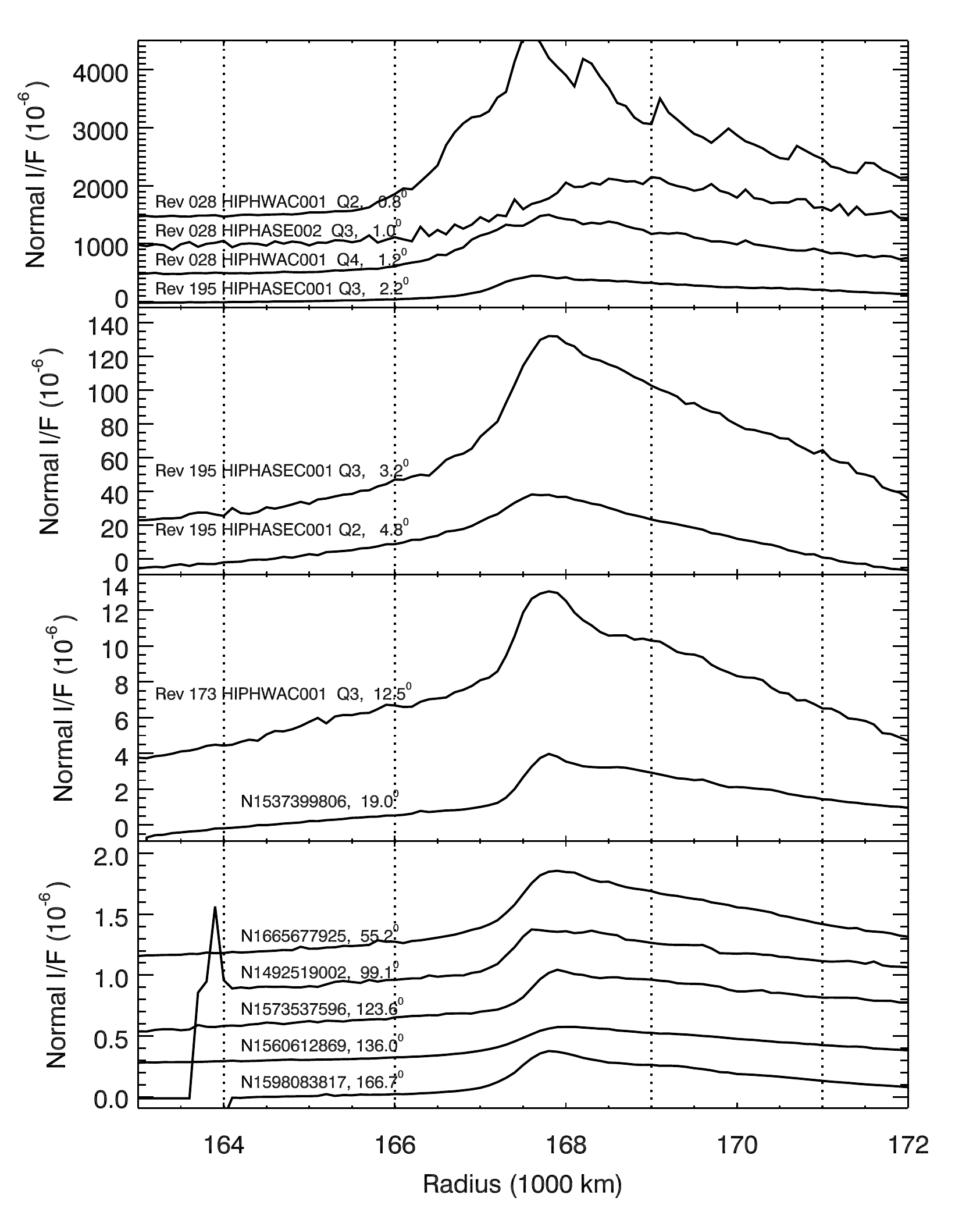}}}
\caption{Representative profiles of the G ring derived from various Cassini images. Each profile is labeled with the Cassini image number and the scattering angle of the ring in that image (for the profiles with scattering angles less than 15$^\circ$, the label indicates the observing sequence and quadrant from which the profile was generated).  Profiles are vertically offset for clarity, and different panels use different vertical scales to cope with the large variations in the ring's brightness with scattering angle. The relatively sharp inner edge of the G ring is visible in all these profiles around 167,500 km. Differences in the shape of this edge are due primarily to differences in image resolution. The vertical dotted lines in each panel indicate the range of radii on either side of the edge that were fit to linear trends to ascertain the brightness contrast across the edge.}
\label{gprof}
\end{figure}

The G ring, on the other hand, is an extended asymmetric feature (see Figure~\ref{gprof}), making it more difficult to quantify in terms of an equivalent width. In particular, we cannot fit the ring's asymmetric profile to a simple shape like a Gaussian or a Lorentzian peak. Also, the outer edge is very diffuse,  gradually fading out against the background E ring over several tens of thousands of kilometers, so it is difficult to isolate the ring signal from background trends in this region. In order to avoid these difficulties, we quantify the brightness of the G ring in terms of the  {\em brightness contrast} across the ring's relatively sharp inner edge. This contrast is computed by fitting linear trends to the brightness-versus-radius curves in two regions on either side of the edge (164,000-166,000 km and 169,000-171,000 km) and extrapolating both those trends to 167,500 km. The difference between these two numbers provides an estimate of the brightness contrast across the edge, and the uncertainty in this value was estimated from the errors on the two linear fits.

\subsection{Selection of high-quality brightness estimates}

The above procedures will only provide sensible estimates of the relevant ring's brightness if they cover a sufficient range of radii, so we visually examined the relevant profiles to ensure that they contained sufficient data to fit the background models for D68 or determine the linear trends on either side of the G ring. These algorithms would also only yield suitable brightness estimates if the data had adequate signal-to-noise, and so additional criteria were used to exclude less reliable measurements.

For the observations at scattering angles greater than 15$^\circ$, where the relevant ring features were subtle, we visually screened the relevant profiles and removed data where spurious peaks from cosmic rays corrupted the data in the vicinity of the relevant ring structure. For the G ring, we also excluded profiles where the inner slope was negative or the outer slope was positive, because such results indicate that the images had complex backgrounds that might contaminate the final calculations. Furthermore, we excluded G-ring images where the fractional error on the contrast was greater than 20\% if the scattering angle was greater than 70$^\circ$, or more than 5\% the contrast if  the scattering angle was smaller than 70$^\circ$. Finally, we excluded the G-ring images that captured the bright arc near the ring's inner edge \citep{hedman2007} because this analysis is concerned with the background G ring.

For observations made at scattering angles less than $15^\circ$, the rings are considerably brighter and so signal-to-noise was less of a concern. Still, we needed to exclude observations where instrumental backgrounds or signals from Saturn's limb contaminated the profiles. For D68, we simply excluded any observation where the background brightness profile did not increase with distance from the planet (indicating a significant background from Saturn's limb) or where there was no peak in the profile $<$ 500 km wide at the expected location for D68. For the G ring, we again required that the inner slope to be positive or the outer slope to be negative, and we excluded G-ring images where the error on the contrast was greater than 5\% the estimated contrast. In this case, we did not deliberately exclude regions containing the G-ring arc because at low scattering angles the arc is a small perturbation on the overall profile.

\section{Results}
\label{results}

\begin{table}
\caption{Normal Equivalent Width estimates for D68 derived from entire images, almost all of which were obtained at scattering angles above 15$^\circ$ (Full Table available in online supplement).}
\label{dtab1}
\begin{tabular}{|c|c|c|c|c|c|c|} \hline
Image Name  & Ephemeris Time &  $\theta$ &  B &   NEW-Int &  NEW-Fit \\
& & (deg) & (deg) & (m) & (m) \\ \hline
N1721658014 & 396235731.719373   &  10.67   &   -9.57   & 6.59   &  5.42 \\
N1547167883 & 221746779.842845    & 17.32   &  25.93   &  3.03   &  2.75 \\
N1547168273 & 221747169.840369    & 17.32   &   25.90  &  2.58   &  2.72 \\
N1547167493 & 221746389.845322    & 17.32    &  25.97  &   2.95   &  2.77 \\ 
N1547167103 & 221745999.847799    & 17.33    &  26.01  &   3.43   &  2.85 \\
N1547166713 & 221745609.850276    &  17.33   &   26.04  &   2.62   &  2.88 \\ 
N1547166323 & 221745219.852752    &  17.33   &   26.08  &   3.05   &  2.95 \\
N1547165933 & 221744829.855229    &  17.33   &   26.12  &   3.11   &  2.90 \\
N1547165543 & 221744439.857706    &  17.33   &   26.15   &  3.02  &   2.92 \\
\hline
\end{tabular}

\end{table}

\begin{table}
\caption{Normal Equivalent Width estimates for D68 derived from observations at scattering angles below 15$^\circ$ (Full Table available in online supplement).}
\label{dtab2}
\begin{tabular}{|c|c|c|c|c|c|c|} \hline
Image Name  & Ephemeris Time  &  $\theta$ &  B &  Longitude &  NEW-Int &  NEW-Fit \\
& & (deg) & (deg) & Range (deg) & (m) & (m) \\ \hline
W1537005177 & 256801306.743551 & 0.56  & 15.42 & 293.41-299.88 & 10048.16 & 15297.32 \\
W1537006505 & 256801339.747224 & 0.61  & 15.40 & 293.37-299.97 & 9060.68 &15362.94 \\
W1537005177 & 336715092.863261 & 0.63  & 15.37 & 286.71-293.29 & 8709.91 &12524.48 \\
W1537006505 & 336719142.834170 & 0.68  & 15.35 & 286.66-293.25 & 7779.63 &12043.77 \\
W1537005177 & 414063384.998397 & 0.69  & 15.31 & 280.00-286.59 & 7818.31 & 10072.35 \\
W1537005177 & 414063701.031543 & 0.73  & 15.25 & 273.29-279.88 & 7119.29  & 9211.29 \\
W1537006505 & 211584138.991393 & 0.74  & 15.29 & 280.06-286.54 & 7002.43 & 9404.19 \\
W1537005177 & 211585466.986816 & 0.76  & 15.20 & 266.71-273.17 & 6876.74 & 9798.16 \\
W1537006505 & 211607287.795406 & 0.78  & 15.23 & 273.34-279.94 & 6584.09 & 8547.55 \\
\hline
\end{tabular}
\end{table}

\begin{table}
\caption{Normal $I/F$ contrast across the G-ring's inner edge derived from observations at scattering angles above 15$^\circ$ (Full Table available in online supplement).}
\label{gtab1}
\begin{tabular}{|c|c|c|c|c|c|} \hline
Image Name   & Ephemeris Time  &  $\theta$ & B & Contrast &  Error \\
& & (deg) & (deg) & (10$^-6$) & (10$^-6$) \\ \hline
N1537372644 & 211951594.636115   &  17.74     &  9.91 &   2.90561  &   0.06985 \\
N1537373650 & 211952600.629690    & 17.78    &   9.90  &  2.85942  &  0.06722 \\
N1537374656 & 211953606.623264   &  17.83   &    9.88  &  2.78850  &  0.07595 \\
N1537375662 & 211954612.616838   &  17.87    &   9.86  &  2.81788  &  0.06576 \\
N1537378680 & 211957630.597561   &  18.00   &    9.81  &  2.81277  &  0.06664 \\
N1537379686 & 211958636.591135   &  18.05   &    9.80  &  2.64955  &  0.06149 \\
N1537380692 & 211959642.584710   &  18.09  &     9.78  &  2.74368  &  0.07522 \\
N1537381698 & 211960648.578284   &  18.14   &    9.76  &  2.87448  &  0.08296 \\
N1537382704 & 211961654.571858   &  18.18   &    9.75  &  2.80607  &  0.07743 \\
N1537383710 & 211962660.565433   &  18.23  &     9.73  &  2.79316  &  0.08281 \\
\hline
\end{tabular}
\end{table}

\begin{table}
\caption{Normal $I/F$ contrast across the G-ring's inner edge derived from observations at scattering angles below 15$^\circ$ (Full Table available in online supplement).}
\label{gtab2}
\begin{tabular}{|c|c|c|c|c|} \hline
Observation   & Quadrant  &  $\theta$ & Contrast &  Error \\
Sequence & & (deg)  & (10$^-6$) & (10$^-6$) \\ \hline
Rev 028 HIPHWAC001 & 2   &  0.80 &    2046.49  &    46.83 \\
Rev 028 HIPHASE002 & 3   &  1.00   & 1265.64   &   35.21 \\
Rev 028 HIPHWAC001 & 4   &  1.20  &   821.38   &   10.79 \\
Rev 028 HIPHASE002 & 1   &  1.20   &  819.20    &  32.94 \\
Rev 195 HIPHASEC001 & 2   &  1.20  &  1124.98  &    39.04 \\
Rev 028 HIPHWAC001  & 4   &  1.40   &  614.81   &    5.45 \\
Rev 028 HIPHWAC001 & 1    & 1.60   &  466.21    &  12.89 \\
Rev 028 HIPHWAC001 & 4   &  1.60   &  443.47    &   5.28 \\
Rev 195 HIPHASEC001 & 4 &    1.60  &   549.82 &      2.04 \\
\hline
\end{tabular}

\end{table}

\begin{table}
\caption{Average Normal Equivalent Width estimates for D68.}
\label{dtab3}
\begin{tabular}{|c|c|c|c|c|c|c|} \hline
$\theta$ Range  &  Ave. $\theta$ &   N & NEW-Int & Error$^a$ &  NEW-Fit & Error$^a$ \\
(deg) & (deg) & & (m) & (m) & (m) & (m) \\ \hline
  0.55   0.60  &     0.56   & 1 &10048.161 & 5024.081 &15297.324 & 7648.662 \\
  0.60   0.65  &     0.62   & 2 & 8885.293 & 3141.426 &13943.709 & 4929.846 \\
  0.65   0.70   &    0.68   & 2 & 7798.970 & 2757.352 &11058.062 & 3909.615 \\
  0.70   0.75   &    0.73  &  2 & 7060.861 & 2496.392 & 9307.736 & 3290.782 \\
  0.75   0.80   &    0.78  &  3 & 6733.740 & 1943.863 & 9697.604 & 2799.457 \\
  0.80   0.85   &    0.82  &  2 & 6281.713 & 2220.921 & 9759.794 & 3450.608 \\
  1.45   1.75   &    1.65  & 12 &  909.391  & 131.259 &  955.829  & 137.962 \\
  1.75   2.00   &    1.85  & 17 &  652.560  &  79.134 &  639.408  &  77.540 \\
  2.00   2.50   &    2.31   & 7  & 255.434  &  48.273  & 296.533   & 56.040 \\
  3.00   4.00   &    3.63  & 22 &   64.945  &   6.923  &  64.655   &  6.892 \\
  4.00   5.00   &    4.54  &  8  &  46.305   &  8.186   & 45.396   &  8.025 \\
  5.50   6.50   &    6.04  &  4  &  17.479   &  4.370   & 18.063    & 4.516 \\
  6.50   7.50   &    6.88  & 18 &   13.987   &  1.648  &  14.279   &  1.683 \\
  7.50   8.50   &    8.10  & 17  &  14.608   &  1.771   & 11.919   &  1.445 \\
  9.00  11.00   &    9.75  & 15   &  7.561   &  0.976   &  7.358    & 0.950 \\
 18.00  20.00  &    18.61  & 33  &   2.558   &  0.223   &  2.550   &  0.222 \\
 20.00  25.00  &    20.58  & 19  &   2.442   &  0.280   &  2.369   &  0.272 \\
 25.00  30.00  &    28.28  & 15  &   1.781   &  0.230   &  1.732   &  0.224 \\
 30.00  35.00   &   32.75  & 84   &  1.623   &  0.089   &  1.543   &  0.084 \\
 35.00  40.00    &  37.38  & 36   &  1.368   &  0.114   &  1.419   &  0.118 \\
 40.00  50.00   &   42.35  & 66   &  1.003   &  0.062   &  1.102  &   0.068 \\
 50.00  60.00 &     55.94  & 38   &  0.627   &  0.051   &  0.739   &  0.060 \\
 60.00  70.00  &    64.94  &  6   &  0.289$^b$   &  0.059    & 0.285$^b$   &  0.058 \\
 70.00  90.00   &   77.04  & 54  &   0.450   &  0.031   &  0.551   &  0.038 \\
 90.00 120.00   &  107.91  & 14  &   0.318   &  0.042  &   0.375    & 0.050 \\
120.00 140.00   &  136.56  & 83  &   0.345  &   0.019  &   0.347   &  0.019 \\
140.00 160.00    & 152.45  & 49  &   0.284  &   0.020   &  0.271  &   0.019 \\
160.00 180.00   &  168.04  &  4   &  0.334  &   0.083   &  0.269 &    0.067 \\
\hline
\end{tabular}

$^a$ Error computed assuming a 50\% uncertainty in each measurement.

$^b$ Measurements questionable (see text).
\end{table}

\begin{table}
\caption{Average Normal $I/F$ contrast across the G-ring's inner edge.}
\label{gtab3}
\begin{tabular}{|c|c|c|c|c|c|c|} \hline
$\theta$ Range  &  Ave. $\theta$ &   N & Contrast &  Error$^a$ &  Scatter$^b$ &  Norm. Error$^c$ \\
(deg) & (deg) & & $10^{-6}$ & $10^{-6}$ & $10^{-6}$ & $10^{-6}$ \\ \hline
  0.60   0.90  &     0.80   & 1  & 2046.485  &   46.831  &   --- &   1023.243 \\
  0.90   1.10  &     1.00   & 1  & 1265.638  &   35.206  &   ---  &   632.819 \\
  1.10   1.30  &     1.20   & 3  &  840.770  &    9.916  &    --- &   242.709 \\
  1.30   1.50  &     1.40   & 1  &  614.805  &    5.447  &    --- &   307.403 \\
  1.50   1.70  &     1.60   & 3  &  534.560  &    1.879  &    --- &   154.314 \\
  1.70   1.90  &     1.80   & 3  &  402.540  &    1.536  &    --- &   116.203 \\
  1.90   2.10  &     2.00   & 5  &  341.047  &    1.380  &  126.893 &    76.260 \\
  2.10   2.30  &     2.20   & 5  &  247.076  &    1.558  &   60.313 &    55.248 \\
  2.30   2.50  &     2.40   & 6  &  190.889  &    1.133  &   36.065 &    38.965 \\
  2.50   2.90  &     2.68   & 8  &  137.794  &    0.717  &   23.698 &    24.359 \\
  2.90   3.50  &     3.20   & 9  &   73.509  &    0.342  &   20.430 &    12.251 \\
  3.50   3.90  &     3.65   & 4  &   41.069  &    0.473  &   14.701 &    10.267 \\
  3.90   4.50  &     4.18   & 8  &   26.370  &    0.332  &    7.590 &     4.662 \\
  4.50   4.90  &     4.72   & 5  &   19.201  &    0.270  &    6.476 &     4.293 \\
  4.90   5.90  &     5.48   & 8  &   14.154  &    0.143  &    4.346 &     2.502 \\
  5.90   7.00  &     6.37   & 6  &   11.195  &    0.112  &    1.922 &     2.285 \\
  7.00   8.00  &     7.50   & 6  &    9.504  &    0.111  &    1.328 &     1.940 \\
  8.00  10.00  &     8.70   & 4  &    7.608  &    0.154  &    0.981 &     1.902 \\
 10.00  14.00  &    12.00   & 3  &    4.952  &    0.072  &    0.000 &     1.429 \\
 17.50  18.50  &    18.12  & 15  &    2.789  &    0.017  &    0.064 &     0.360 \\
 18.50  19.50  &    18.92  & 18  &    2.696  &    0.014  &    0.123 &     0.318 \\
 19.50  20.50  &    19.95  & 16  &    2.484  &    0.015  &    0.084 &     0.310 \\
 45.00  54.00  &    50.86  &  9  &    0.721  &    0.008  &    0.035 &     0.120 \\
 54.00  65.00  &    56.46  & 22  &    0.551  &    0.002  &    0.033 &     0.059 \\
 70.00  90.00  &    80.41  &  4  &    0.447  &    0.032  &    0.024 &     0.112 \\
 90.00 110.00  &    99.09  &  1  &    0.371  &    0.017  &   --- &    0.185  \\
110.00 130.00  &   121.59  & 26  &    0.371  &    0.003  &    0.057 &     0.036 \\
130.00 150.00  &   136.47  &  9  &    0.271  &    0.001  &    0.030 &     0.045 \\
150.00 165.00  &   159.02  & 10  &    0.286  &    0.008  &    0.046 &     0.045 \\
165.00 180.00  &   166.09  &  5  &    0.311  &    0.003  &    0.016 &     0.070 \\
\hline
\end{tabular}

$^a$ Error computed by propagating errors on individual measurements.

$^b$ $rms$ scatter of measurements.

$^c$ Error computed assuming a 50\% uncertainty in each measurement.

\end{table}

\begin{figure}
\centerline{\resizebox{3in}{!}{\includegraphics{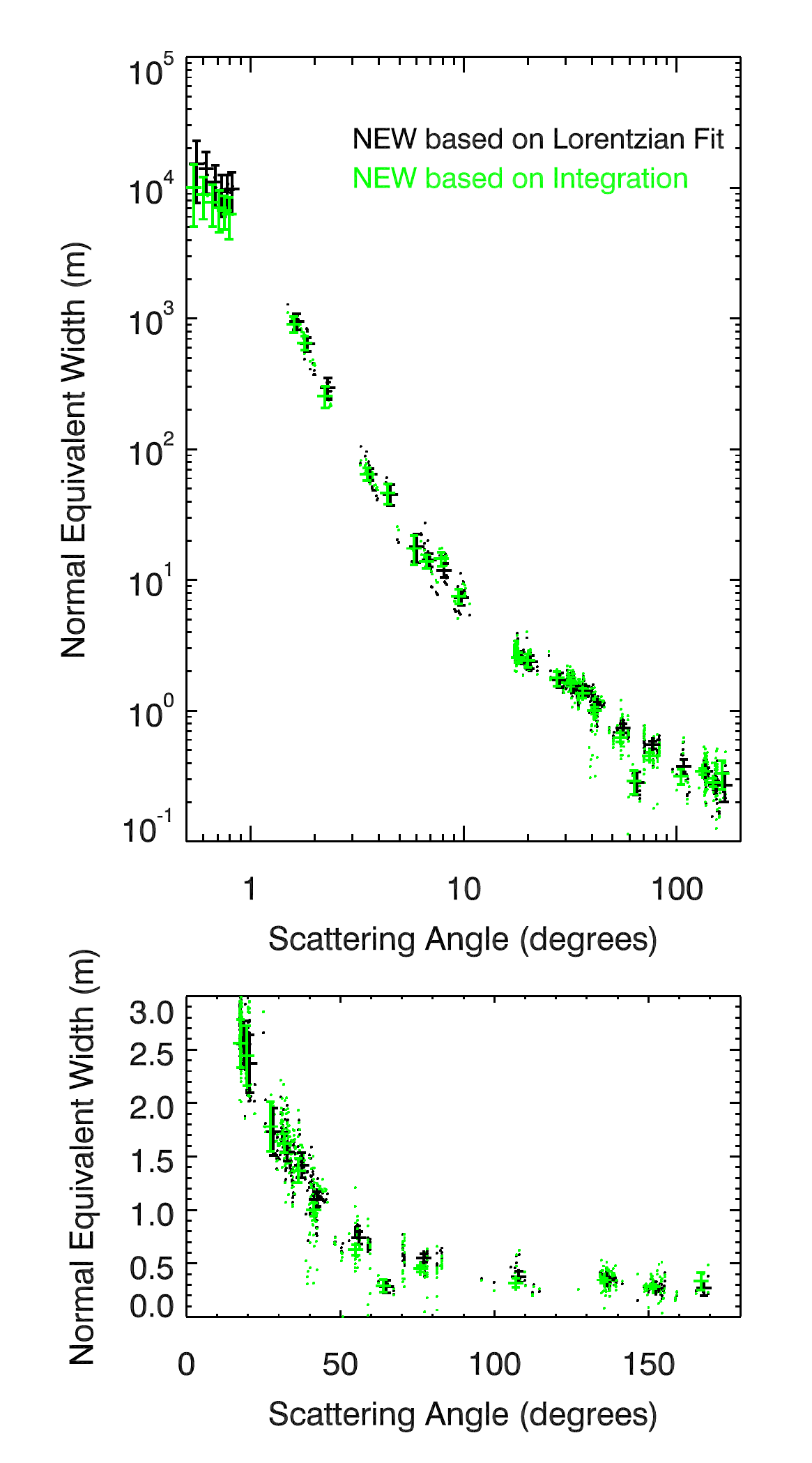}}}
\caption{Phase curve of the D68 ringlet. The black and green points correspond to the two different methods of estimating the integrated brightness of this feature. Individual brightness estimates are shown as points, while the error bars correspond to the binned averages and uncertainties, computed assuming each data point has an intrinsic uncertainty of 50\%.}
\label{d68phase}
\end{figure}

\begin{figure}
\centerline{\resizebox{3in}{!}{\includegraphics{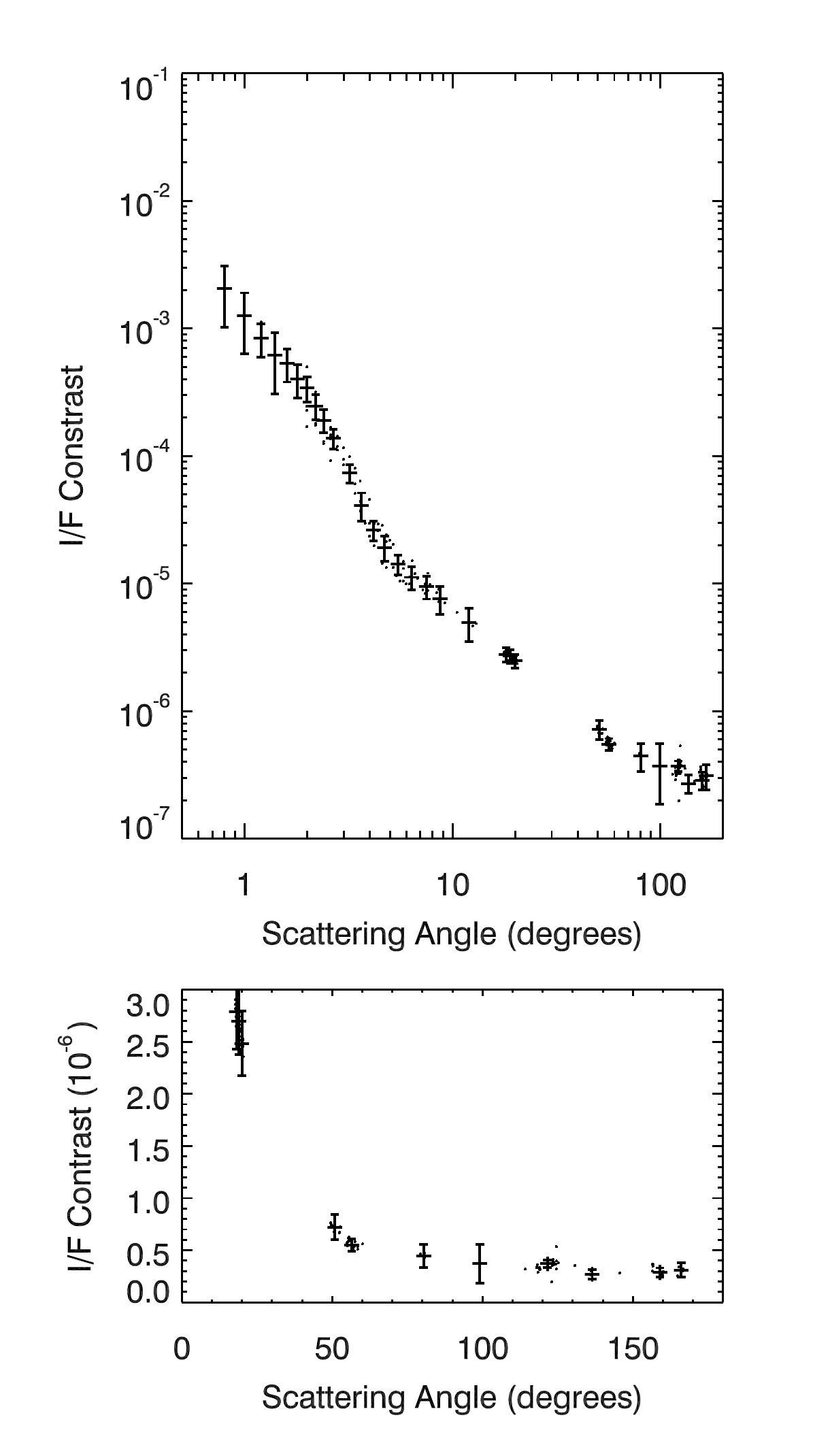}}}
\caption{Phase curve of the brightness contrast across the G-ring's inner edge. Individual brightness estimates are shown as points, while the error bars correspond to the binned averages and uncertainties, computed assuming each data point has an intrinsic uncertainty of 50\%.}
\label{gphase}
\end{figure}

\begin{figure}
\centerline{\resizebox{3in}{!}{\includegraphics{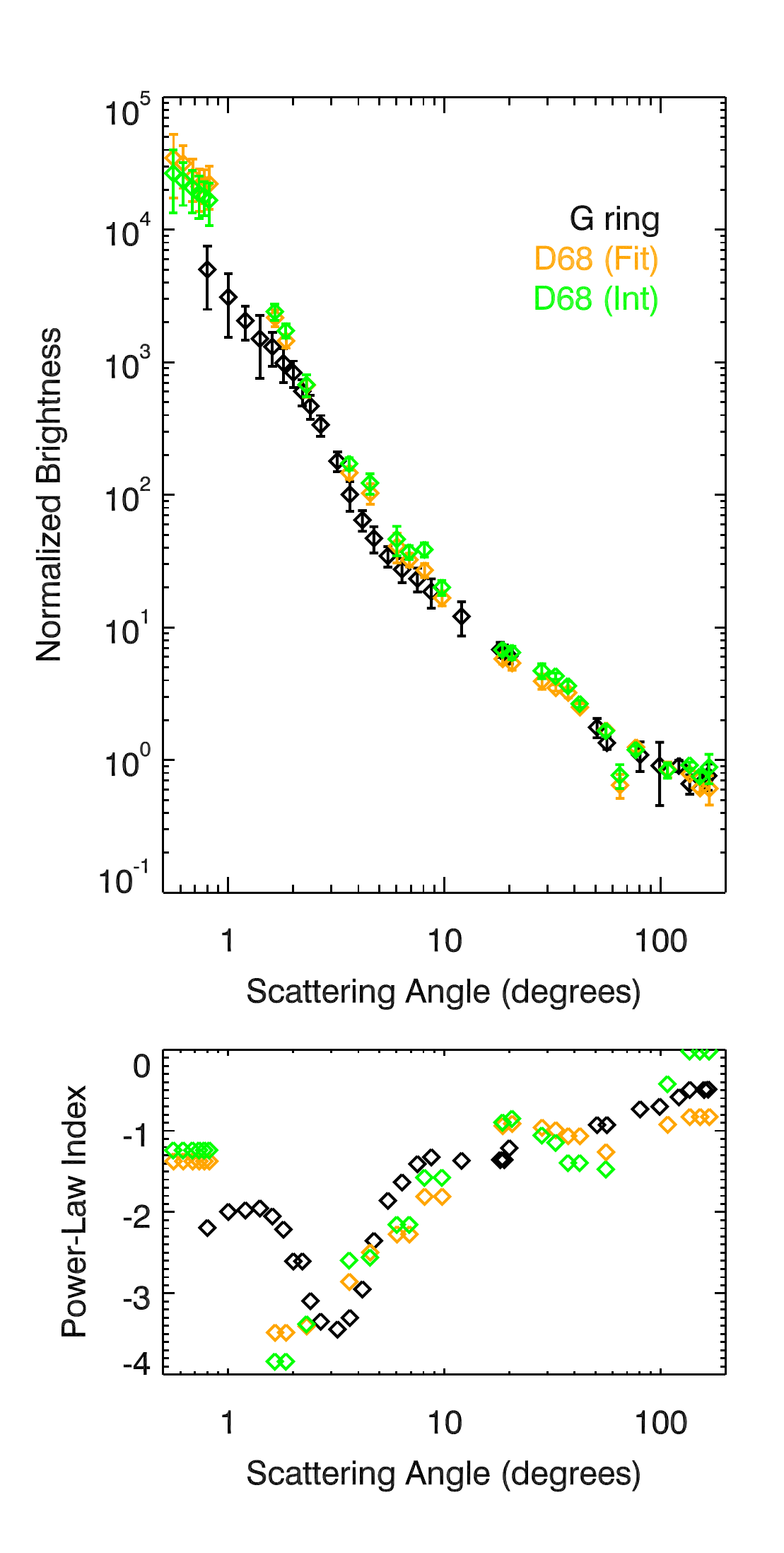}}}
\caption{Comparing the SPFs for D68 and the G ring. The upper plot displays the normalized phase curves of the two ring features, showing that they have similar, but not identical shapes. The bottom panel shows the power-law index of the SPFs as a function of scattering angle.}
\label{dgphase}
\end{figure}

The above procedures yielded 751 estimates of D68's Normal Equivalent Width and 225 estimates of the brightness contrast across the G-ring's inner edge. These estimates, along with the relevant geometrical parameters, are provided in Tables~\ref{dtab1}-\ref{gtab2}. These individual estimates of the rings' brightness are plotted as a function of phase angle as the small data points in Figures~\ref{d68phase} and~\ref{gphase}.  For both the G ring and the D68 Fit measurements, the scatter of these data points around the mean trend is less than 50\% (the D68 Int measurements having somewhat larger scatter). This dispersion, while small enough for the purposes of this analysis, is larger than the expected statistical errors in these parameters based on the relevant fits, indicating that systematic uncertainties dominate the scatter. For D68, the observed scatter is comparable to the known longitudinal asymmetries in this ring feature \citep{hedman2014}, so real brightness variations within the ringlet are probably responsible for most of the observed dispersion in the D68 Fit brightness estimates. The larger scatter in the D68 Int measurements probably arise because this method is more sensitive to structures in the background under the ring.  Similarly, the scatter in the G-ring data is probably associated with systematic errors in the brightness contrast estimates due to varying background signals in the images. 

To further facilitate the analysis of these brightness data, we define a series of bins in scattering angle and compute the average brightness of the relevant ring feature in each bin, along with the corresponding uncertainty in this average brightness level. This uncertainty is estimated by assuming each data point has an independent systematic error of 50\%. This should provide a conservative estimate of the uncertainty in the SPF because the $rms$ dispersion of the measurements within each bin (which is predominantly due to the systematic phenomena discussed above) rarely exceeds 50\%. These estimates of the rings' brightness are provided Tables~\ref{dtab3} and~\ref{gtab3} and are shown as points with error bars in Figures~\ref{d68phase} and~\ref{gphase}. Both sets of brightness data span a broad range of scattering angles between 0.5$^\circ$ and 170$^\circ$, and the brightness decreases over four orders of magnitude with increasing scattering angle. The one notable exception to this generally decreasing trend are the six estimates of the D68's brightness between scattering angles of 60$^\circ$  and 70$^\circ$, which fall about a factor of two below the measurements at slightly higher and lower scattering angles. Thus far, we have been unable to identify any instrumental or analytical phenomenon that could explain these discrepant results, but such a narrow dip in the phase function seems unphysical, and so we will treat these measurements with suspicion for the remainder of this investigation. 

Our brightness estimates of D68 and the G ring are reasonably consistent with the limited Voyager observations of these features \citep{showalter1993, showalter1996, hedman2007d}. The observed trends 
in the brightness of these ring features are also similar to the previously published phase curves of Jupiter's main ring and Saturn's F ring \citep{throop2004, french2012}. All these rings contain a strong forward-scattering peak together with a relatively flat phase function for scattering angles above about 60$^\circ$. However, the extensive data for D68 and the G ring compiled here now enable us to examine these brightness variations in more detail. 

The brightness profiles shown in Figures~\ref{d68phase} and~\ref{gphase} are proportional to the average scattering phase function of these rings, with the proportionality constant set by the optical depth of the ring and bond albedo of the ring particles. Ideally, the SPF is normalized so that the integral of the function over all solid angles is either unity or $4\pi$. Unfortunately, even though our observations cover almost the entire range of possible scattering angles, we cannot compute the relevant integral reliably because the brightness increases so rapidly in the core of the forward-scattering peak. Even if the rings' brightness scales like $1/\theta$ between 0$^\circ$ and 0.5$^\circ$, that last half-degree would still contribute an order of magnitude more to the integral than all other solid angles. The derived integral is therefore extremely sensitive to the assumed shape of the last 0.5$^\circ$ of the forward scattering peak.

Since we cannot robustly determine the integral under the phase curve, we will instead normalize the SPFs to be unity where the scattering angle is 90$^\circ$. These normalized SPFs are shown in Figure~\ref{dgphase}, which reveals that the SPFs of the G ring and D68 are remarkably similar to each other.  Another way to look at these data is shown in the bottom panel of Figure~\ref{dgphase}, where we plot the local value of the power-law index in the SPF as a function of scattering angle. To generate this plot, we fit the SPF in the vicinity of each data point\footnote{Each fit includes all data obtained at scattering angles within $\pm50$\% of the selected data point, excluding the discrepant point in D68's SPF around $\theta=65^\circ$.} to a simple power-law function (i.e. SPF$\propto\theta^{Q}$). The plot shows how the index $Q$ varies with scattering angle and reveals that both SPFs can be divided into roughly four regimes:
\begin{itemize}
\item
{\bf Scattering angles above 50$^\circ$}, where the power law index is close to zero, indicating that the brightness of both features does not depend much on scattering angle. 
\item
{\bf Scattering angles between 10$^\circ$ and 50$^\circ$}, where the ring's brightness scales inversely proportionally with scattering angle.
\item
{\bf Scattering angles between 2$^\circ$ and 10$^\circ$}, where the scattering function becomes dramatically steeper, with a power-law index approaching -3.5.
\item
{\bf Scattering angles less than 2$^\circ$}, where the scattering angle becomes shallow again.
\end{itemize}

\section{Fits to the measured scattering phase functions}
\label{fits}

In order to further quantify the trends observed in the above data and clarify their implications, we compare the measurements to various models. For these studies, we use the normalized, binned data displayed in Figure~\ref{dgphase} and only consider the estimates of D68's brightness derived using the Fit method. We also exclude the anomalously low D68 measurement at scattering angles around 65$^\circ$. 

First, we use Henyey-Greenstein functions to derive simple analytical expressions that reproduce the relevant observations. Next, we compare the ring data to laboratory measurements of real particle populations. These comparisons demonstrate that the basic shapes of our SPFs are reasonable, but their detailed form is different from currently-available laboratory samples. Finally, we consider physically-motivated models based on Fraunhofer and Mie theories in an attempt to gain some insights into the particle size distributions of these rings. 

\subsection{Henyey-Greenstein functions}

\begin{deluxetable}{lllllllll}
\tablewidth{0pt}
\footnotesize
\tablecaption{Best fit HG functions \label{HG_fits_table}}
\tablehead{
\colhead{Ring} & \colhead{$\chi^2/\nu$} & \colhead{$p_>^a$} & \colhead{$g_1$} & \colhead{$w_1$} & \colhead{$g_2$} & \colhead{$w_2$} & \colhead{$g_3$} & \colhead{$w_3$}   \\
\vspace{-0.1in}
}
\startdata
\vspace{+0.05in}
G 	  & $12.3$ & $<10^{-16}$ & $0.525^{+0.475}_{-0.525}$  & $1.0$   \\
\vspace{+0.05in}
D68      & $22.8$ & $<10^{-16}$ &  $0.393^{+0.430}_{-0.384}$  & $1.0$  \\
\vspace{+0.05in}
G   & $2.5$   & 0.00045 & $0.985^{+0.015}_{-0.045}$ & $0.447^{+0.342}_{-0.221}$ & $0.345^{+0.255}_{-0.305}$  &  $0.553^{+0.221}_{-0.342}$   \\
\vspace{+0.05in}
D68  & $4.6$   & $1.2\times10^{-12}$ &  $0.995^{+0.005}_{-0.055}$ &  $0.779^{+0.07}_{-0.443}$ & $0.325^{+0.165}_{-0.155}$ & $0.221^{+0.443}_{-0.07}$   \\
\vspace{+0.05in}
G   & $0.71$  & 0.84  & $0.995^{+0.005}_{-0.015}$ & $0.643^{+0.076}_{-0.306}$ & $0.665^{+0.085}_{-0.185}$ & $0.176^{+0.266}_{-0.106}$ & $0.035^{+0.225}_{-0.035}$ & $0.181^{+0.291}_{-0.101}$ \\
\vspace{+0.05in}
D68  &  $1.5$  & 0.15   & $0.995^{+0.005}_{-0.015}$ & $0.754^{+0.055}_{-0.247}$ & $0.585^{+0.165}_{-0.165}$ & $0.151^{+0.166}_{-0.096}$ & $0.005^{+0.255}_{-0.005}$ & $0.095^{+0.146}_{-0.045}$\\
\enddata

$^a$ Probability that the $\chi^2/\nu$ parameter for the best-fit model would be at least as large as its observed value if the selected model were correct.
\end{deluxetable}

\afterpage{\clearpage}

\begin{figure}[H]
\centerline{\resizebox{5in}{!}{\includegraphics{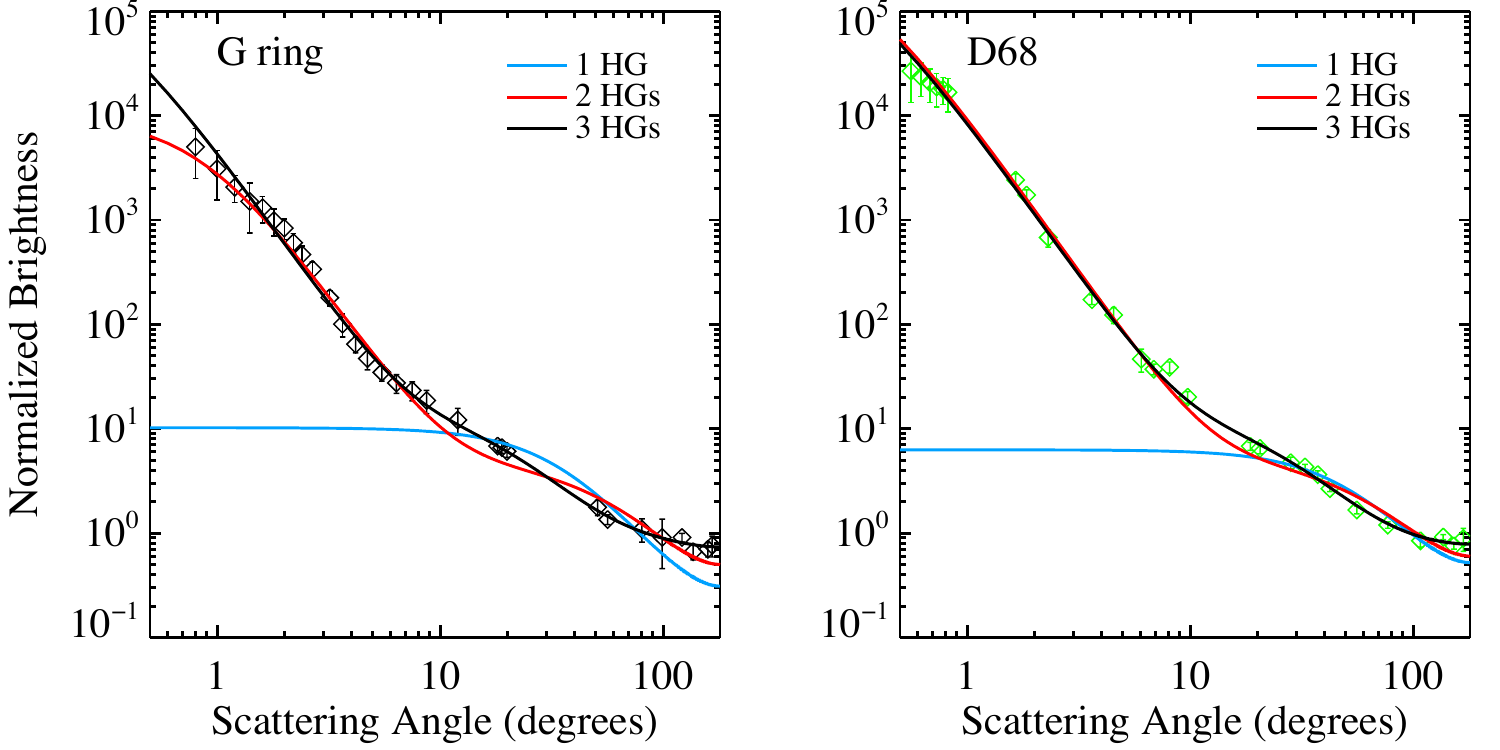}}}
\caption{Fits to the normalized brightness for the G ring (left) and D68 (right), using 1-, 2-, and 3-component HG fits.  Table \ref{HG_fits_table} lists the best fit parameters.}
\label{HG_fits}
\end{figure}

The Henyey-Greenstein (HG) function is the function most commonly used to describe the SPF of debris disks, and can be expressed as follows:
\begin{equation}
	p\!\left(g,\theta\right) = \frac{1}{4\pi}\frac{1-g^2}{[\,1+g^2-2g\cos{\theta}\,]^{3/2}},
\end{equation}
where $\theta$ is the scattering angle and $g$ is the HG asymmetry parameter,which ranges from -1 for perfect backscattering to 1 for perfect forward scattering \citep{henyey1941}.  We fit the normalized brightnesses using a single HG function, examining all possible values of $g$ in steps of $0.001$ and varying the amplitude of the fit to minimize the $\chi^2$ metric.  The best fits, shown as blue lines in Figure \ref{HG_fits}, correspond to $g=0.525$ for the G ring and $g=0.393$ for D68.  However, these fits have $\chi^2/\nu = 12.3$ and $22.8$, respectively ( $\nu=$ degrees of freedom for the fit), and the probability of the best-fit $\chi^2$ values being at least this large just due to random chance is extremely low (i.e. $p_> <10^{-16}$). A single HG function is therefore a very poor fit to the measured SPFs.

In an attempt to achieve an acceptable match to the data (i.e. $\chi^2/\nu \sim 1$ and $p_> \sim 0.5$), we fit the measured SPFs using a 2-component HG function of the form
\begin{equation}
	p_2\!\left(g_1,w_1,g_2,\theta\right) = w_1\; p\!\left(g_1,\theta\right) + w_2\; p\!\left(g_2,\theta\right),
\end{equation}
where $w_2=(1-w_1)$ and $w_1$ ranges from 0 to 1.  We examined all possible combinations of $g_1$ and $g_2$ using step sizes of $0.01$, and all possible values of $w_1$ in steps of $0.005$.  The best fit parameters are listed in Table \ref{HG_fits_table} and result in $\chi^2/\nu = 2.5$ and $4.6$ for the G ring and D68, respectively.  These best fits are shown as red lines in Figure \ref{HG_fits}. While these fits are much better than the single HG solutions, the probabilities that the $\chi^2$ statistics would be this large are still below 0.001, indicating that the data deviate significantly from this model.

Finally, we examined 3-component HG functions of the form
\begin{equation}
	\label{hg3comp_equation}
	p_3\!\left(g_1,w_1,g_2,w_2,g_3\theta\right) = w_1\; p\!\left(g_1,\theta\right) + w_2 \; p\!\left(g_2,\theta\right)+ w_3 \; p\!\left(g_3,\theta\right),
\end{equation}
with the requirements $w_3 = (1-w_1-w_2)$ and $w_3 \ge 0$.  To make a brute force $\chi^2$ calculation more tractable, we restricted $0.5 < g_1 < 1$, $0.25 < g_2 < 0.75$, and $0 < g_3 < 0.5$ using step sizes of $0.01$.  Given the results of the 2-component fit, and the lack of a back-scattering feature in the measured SPF, we don't expect these limitations to impact the best derived fit.  We examined all possible, valid combinations of $w_1$ and $w_2$ in steps of $0.005$.  The best fit parameters are listed in Table \ref{HG_fits_table}.  The 3-component fits have significantly better $\chi^2/\nu$ statistics, with values of $0.71$ and $1.5$  (and corresponding $p_<$ values of 0.84 and 0.15 ) for the G ring and D68, respectively.  These best fit 3-component functions are shown as black lines in Figure \ref{HG_fits} and reproduce the data well. Hence, If one's goal is to quickly approximate the SPFs of the G ring or D68, we recommend using these three-component fits. However, we do not regard the parameters of these fits as having any specific physical significance because a HG function only provides a rough approximation of any given system's light-scattering properties.

\subsection{Empirical comparisons}

While the HG functions can  reproduce the observed brightness data, the HG formalism provides little information about the physical properties of the material in these rings. Insights into the rings' particle properties can potentially be obtained by comparing our data to the Amsterdam Light Scattering Database \citep{munoz2012}.  This database contains high-precision laboratory measurements of scattering phase functions for particles with a wide variety of compositions, including liquid water droplets, clays, olivine, and fluffy aggregates of synthetic circumstellar dust.  Each sample has an estimated size distribution, with effective grain radii typically falling between $0.1$ and $10$ $\mu$m.  We included only the 36 non-hydrosol samples that were measured at a wavelength of $0.63$ $\mu$m.

To compare these laboratory measurements with our astronomical SPFs, we interpolated the laboratory SPF measurements and their fractional uncertainties in log-log space.  We added the uncertainty of both data sets in quadrature when calculating a best fit.  Our comparison of the data sets was limited to $\theta>3^{\circ}$ due to the limitations of the Amsterdam Light Scattering Database experiment \citep{munoz2012}.  Note that we are unable to adjust the grain size distribution of the laboratory measurements, and so better fits may be possible with grain size distributions that are not currently available.

\begin{figure}[H]
\centerline{\resizebox{5in}{!}{\includegraphics{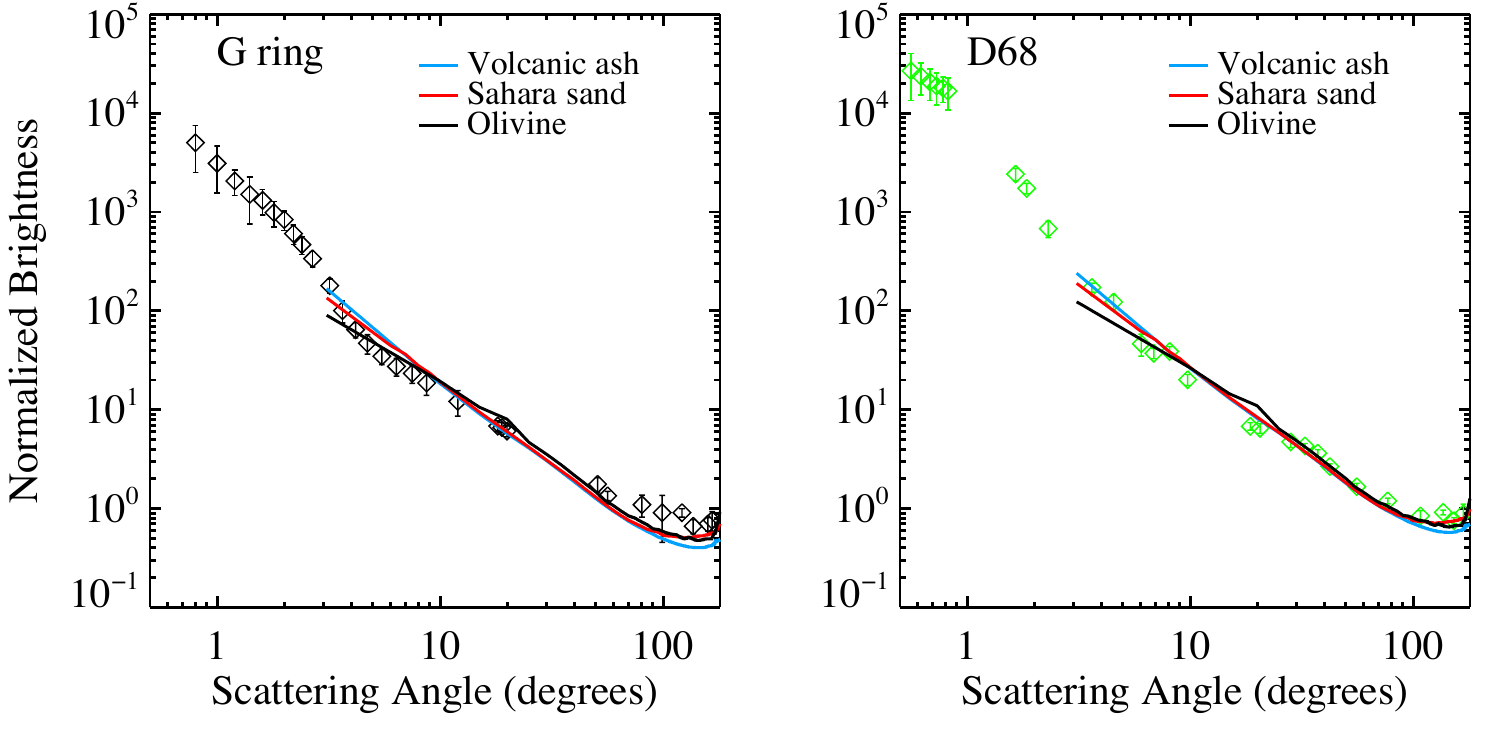}}}
\caption{Fits to the measured SPF using the Amsterdam Light Scattering Database.  Fits are limited to $\theta>3^{\circ}$.  SPFs for large grain olivine, Sahara sand (Libyan sample shown), and volcanic ash (Spurr Gunsight sample shown) generally fit best.  }
\label{empirical_fits}
\end{figure}

We were unable to achieve $\chi^2/\nu < 3.4$ for any of the compositions in the database. In particular, most of the 7 different samples of fluffy synthetic circumstellar dust aggregates were poor fits to our data. Large grain olivine samples ($s_{\rm eff}>4$ $\mu$m), Sahara sand samples, and volcanic ash samples fit best with $\chi^2/\nu \lesssim 7$ for the G ring and $\chi^2/\nu \lesssim 8$ for D68. Figure~\ref{empirical_fits} shows three examples of these SPFs overlaid on our data. While the matches are not particularly good, these laboratory SPFs do at least demonstrate that the overall shape of the rings' SPFs are not unreasonable for $\theta>3^\circ$. They also suggest that the rings' SPFs are not particularly sensitive to the particles' compositions.

\subsection{Fraunhofer diffraction}

In lieu of more extensive laboratory measurements, we can fit our data to theoretical models based on Mie theory and Fraunhofer diffraction. Ideally, such theories would allow us to translate features in the observed SPF into information about the particle size distribution. However, in practice the approximations inherent in these theories and the large parameter space of possible particle size distributions complicates efforts to find unique and well-defined solutions. Thus the following analyses should be considered initial explorations that can provide some insights into these rings' particle size distributions and guide future efforts to model the photometric properties of these systems.

Fraunhofer diffraction theory provides the simplest light-scattering model that can translate a ring's observed SPF into information about its particle size distribution. This model assumes that the light scattered by each particle in the ring has a forward-scattering peak with a characteristic width set by the ratio of the particle's radius $s$ to the wavelength of the scattered light $\lambda$. Strictly speaking, classical Fraunhofer diffraction is only valid in the far-field limit of small opaque particles, but at very small scattering angles  ($\theta \ll \lambda/s$) Fraunhofer theory can provide a useful approximation of light-scattering properties for even the ice-rich particles found in Saturn's rings \citep{hedman2009}. In this limit, the brightness of the ring material should be given by the following expression:
\begin{equation}
\mu I/F = \mathcal{C}\int \frac{\pi s^2}{\sin^2\theta}\frac{dN}{ds}J_1^2(k s\sin\theta) ds,
\end{equation}
where $\mathcal{C}$ is a constant numerical coefficient, $k=2\pi/\lambda$ is the wavenumber of the observed radiation and $dN/ds$ is the number of particles in a small range of particle size $ds$. Note that if the differential size distribution is a pure power law (i.e. $dN/ds \propto s^{q}$), then this expression implies that the combined light from all the particles would be a power-law function of scattering angle (i.e. the observed brightness is $\propto \theta^{-(q+5)}$). Note that a shallower particle size distribution (i.e. less negative $q$) will produce a steeper SPF (i.e. more positive $q+5$). Furthermore, since the square of the Bessel function $J_1^2(x)$ has a peak at $x \simeq 2$, the measured brightness at a given scattering angle $\theta$ will be predominantly due to particles with a size $s \simeq \pi^{-1}\lambda/\theta$ (assuming $q$ is close to $-3$).

The observed SPFs for the G ring and D68 do not follow a simple power law for small scattering angles. Instead, the power-law index of the scattering function becomes dramatically less steep when the scattering angles falls below 2$^\circ$ (see Figure~\ref{dgphase}). This suggests that the particle size distribution does not follow a perfect power law. More specifically, the transition to a lower phase slope at smaller scattering angles implies that particle size distribution becomes steeper (or is cut off) above some critical size. In this situation, we can define an effective particle size $s_{\rm eff}$, which corresponds to the average effective size of the particles in the ring. We can even estimate this parameter by fitting the observed SPF to a Fraunhofer model of the scattering function for a single particle, because the width of the forward scattering peak of a particle with size $s_{\rm eff}$ will be close to the width of the forward scattering peak of a particle size distribution with that effective average size.


We considered a range of possible effective grain sizes from $0.1$--$100$ $\mu$m.  To satisfy the Fraunhofer criterion of small scattering angles, we selected only the data at $\theta < 0.1\lambda/s_{\rm eff}$ for each effective grain size.  We fit this portion of the forward scattering peak with an Airy function (appropriate for spherical grains) and considered only those fits that had at least 1 degree of freedom.  We found a best fit effective grain size of $s_{\rm eff}=2.2^{+0.3}_{-0.7}$ $\mu$m for the G ring, with $\chi^2/\nu = 1.8$ ($p_>=0.40$), and a best fit effective grain size for D68 of $s_{\rm eff}=2.5^{+0.2}_{-0.3}$ $\mu$m, with $\chi^2/\nu = 0.28$ ($p_>=0.96$).  The high probabilities for the $\chi^2$ to exceed the observed values suggests that the uncertainties on the data points in the forward-scattering peak may be overestimated. When relaxing the small angle constraint to $\theta < 0.25 \lambda/s$, best fit grain sizes roughly doubled while $\chi^2/\nu$ remain relatively unchanged.  
Thus our estimate of $s_{\rm eff}$ is likely less certain than the above error bars suggest, and so we can only conclude that the typical particle size in this ring is probably of order a few microns.
 
\subsection{Mie theory}

Mie theory provides exact analytical expressions for the amount of light scattered in all directions by perfect dielectric spheres, and it is commonly used to constrain the composition of debris disks by modeling the disk thermal emission in the infrared, along with the optical and near-IR photometry \citep[e.g.][]{li1998, lebreton2012, rodigas2015}.  
However, Mie theory does not accurately predict the SPF for non-spherical particles \citep[e.g.][]{pollack1980}. The data shown in Figure~\ref{dgphase} provide new opportunities to examine the applicability of Mie theory to real systems of fine debris.
	
\citet{rodigas2015} calculated the scattering phase functions for 8407 unique combinations of materials, stemming from 19 ``root" data sets containing the complex indices of refraction as a function of wavelength.  Here, we fit the measured SPFs using a subset of these materials, requiring that each material contain some fraction of water ice, which is known to be prevalent in Saturn's rings \citep[][and references therein]{cuzzi2009}.  Each ``root" water ice data set was mixed with other combinations of materials according to the rules of \citet{rodigas2015}, resulting in 19 porous water mixtures, 144 2-component mixtures, and 720 porous 2-component mixtures per water ice data set.  We examined three different ``root" water ice data sets, for a total of 2652 mixtures.

We calculated the scattering efficiency and SPF of each mixture at $0.63$ $\mu$m using Mie theory and Bruggeman effective medium theory as described in \citet{rodigas2015}.  We calculated the SPF from 0--$180^{\circ}$ in $0.5^{\circ}$ increments for 50 values of grain size logarithmically spaced from $0.1$--$500$ $\mu$m.  We assumed a power law grain size distribution of the form
\begin{equation}
\frac{dN}{ds} \propto s^{q},
\end{equation}
between the sizes $s_{\rm max}$ and $s_{\rm min}$ and calculated the size-integrated SPF by weighting each grain size by the size distribution power law, grain cross section, and scattering efficiency calculated by Mie theory.  We investigated 51 values of $q$ ranging from $-7$ to $-2$.  We truncated the size distribution at minimum and maximum grain sizes that spanned the full range of investigated sizes.  We interpolated each model SPF in log-log space for comparison with the measured values and calculated a reduced $\chi^2$ value for each mixture.  

\begin{figure}[tb]
\centerline{\resizebox{5in}{!}{\includegraphics{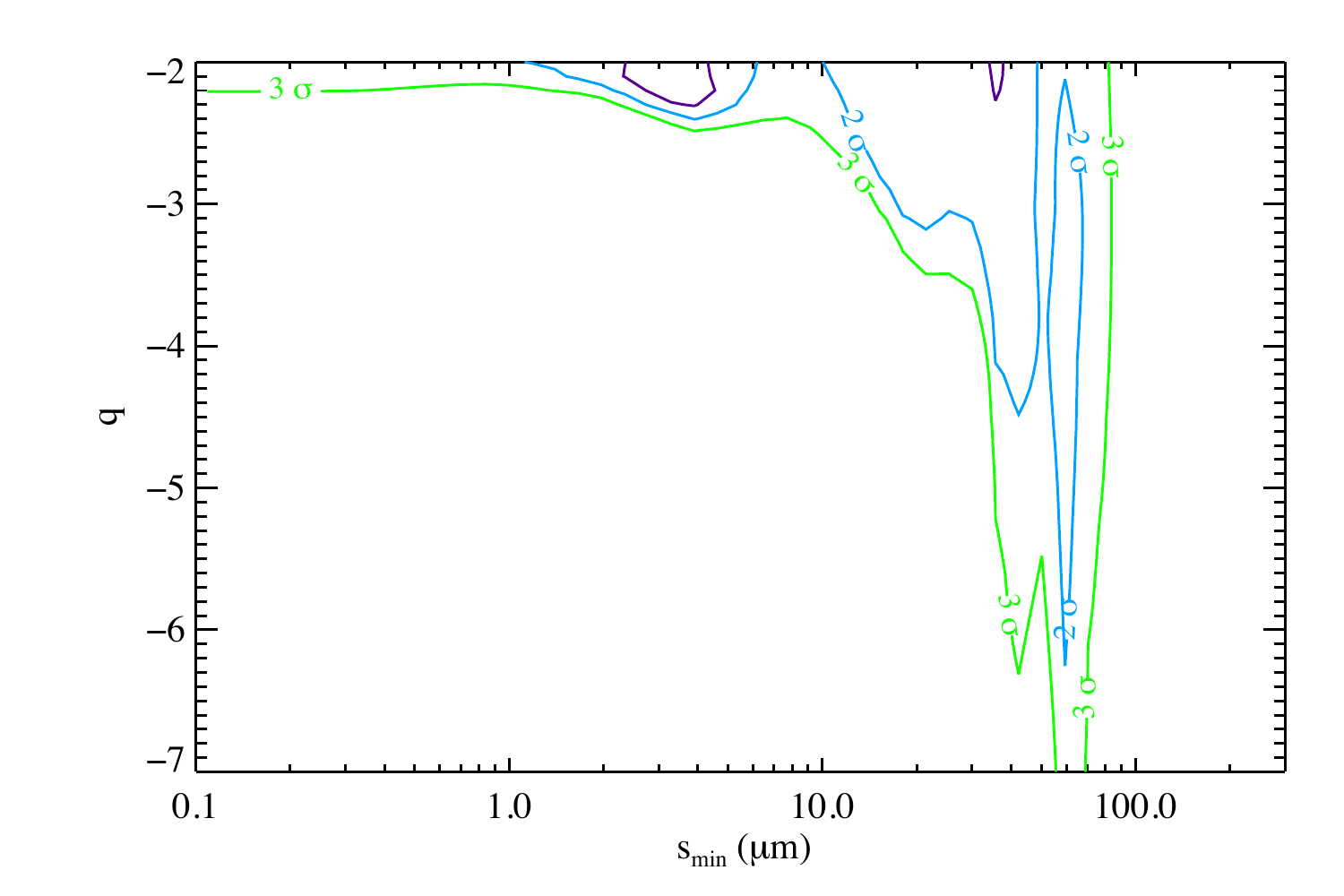}}}
\caption{Projected $\Delta\chi^2$ contours for Mie theory fits to the G ring, labeled by their 2 parameter joint confidence levels.}
\label{mie_theory_fits_chi2contours_G}
\end{figure}

The G ring was best fit by a shallow size distribution ($q=-2.2$) of grains ranging from 4 to 460 $\mu$m, with $\chi^2/\nu=0.7$.  The single best fit composition was a mixture of 10\% water ice and 90\% olivine by volume, with an additional porosity of 10\%. However, this best fit is not unique.  Figure \ref{mie_theory_fits_chi2contours_G} shows the projected $\Delta\chi^2$ contours as a function of $s_{\rm min}$ and $q$.  To create this plot, we projected the minimum $\chi^2$ from all other parameters.  The three contours, labeled as 1, 2, and 3$\sigma$, correspond to the appropriate 2 parameter joint confidence levels for $\Delta\chi^2 = 2.3$, $6.17$, and $14.2$, respectively.  Two 1$\sigma$ ``islands" exist at the top of the plot.  The leftmost of these islands contains shallow size distributions with minimum grain sizes $\sim$few $\mu$m, and includes the best fit.  A total of 221 unique mixtures with a wide range of compositions and porosities are included in this portion of phase space, with water ice volumetric fractions typically $<30\%$.  This suggests that a flat size distribution can reproduce the observations well, regardless of the details of the scattering and optical properties.  The rightmost 1$\sigma$ island requires larger minimum grain sizes $\sim40$ $\mu$m, and includes only three compositions, all of which are roughly an 80\% mixture of iron-rich olivine or iron-rich orthopyroxene mixed with 20\% water ice.

\begin{figure}[tb]
\centerline{\resizebox{3in}{!}{\includegraphics{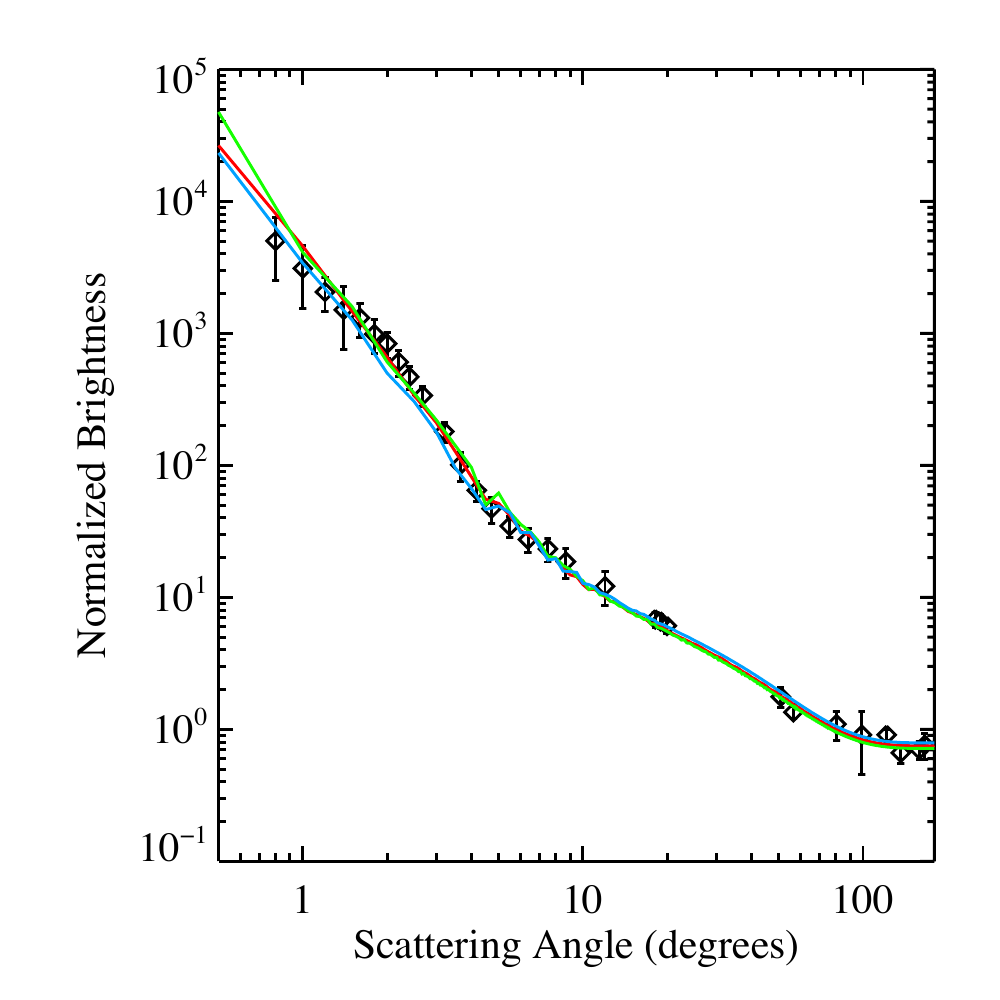}}}
\caption{Three sample Mie theory fits to the G ring SPF, all of which fall within the 2$\sigma$ contour.}
\label{mie_theory_fits_G}
\end{figure}

Figure \ref{mie_theory_fits_G} shows the SPFs for three sample fits.  The red curve shows the model SPF for the best fit.  The green curve corresponds to an $s_{\rm min} = 18$ $\mu$m and $q=-2.8$ fit, which falls within the 2$\sigma$ contours shown in Figure \ref{mie_theory_fits_chi2contours_G}.  The blue curve corresponds to an $s_{\rm min} = 35$ $\mu$m and $q = -3.2$ fit.  All three of these fits predict that the forward-scattering peak continues to rise steeply, such that the SPF is four orders of magnitude brighter at $\theta=0^{\circ}$ than at $\theta \approx 1^{\circ}$. 

\begin{figure}[htb]
\centerline{\resizebox{5in}{!}{\includegraphics{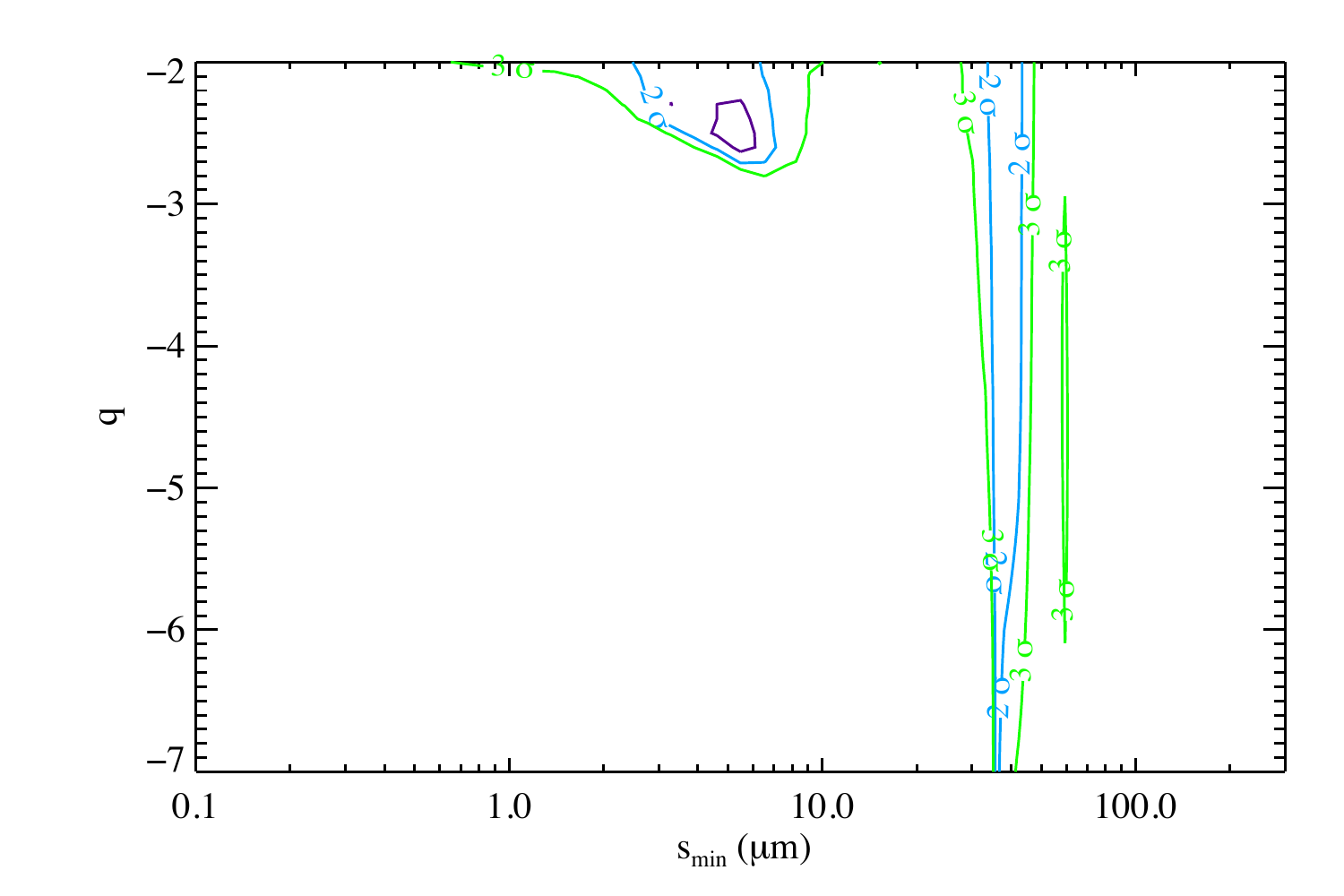}}}
\caption{Projected $\Delta\chi^2$ contours for Mie theory fits to the D ring.}
\label{mie_theory_fits_chi2contours_D}
\end{figure}

Similar to the G ring, the D68 ringlet was best fit by a shallow size distribution ($q=-2.5$) of grains ranging from 5 to 390 $\mu$m, with $\chi^2/\nu=0.9$.  The single best fit composition was a mixture of 10\% water ice and 90\% iron-rich orthopyroxene by volume, with an additional porosity of 10\%.   Figure \ref{mie_theory_fits_chi2contours_D} shows the $\Delta\chi^2$ contours for the D68 ringlet.  Here the 1$\sigma$ best fits are well isolated around the overall best fit and 2 distinct regions of parameter space are included within 3$\sigma$.  Like the G ring fits, the SPF of the D68 ringlet is best fit by either a shallow size distribution with minimum grain size $\sim$few $\mu$m, or a size distribution with minimum grain size $\sim$few tens of $\mu$m.  Within the 1$\sigma$ limit, a wide range of compositions and porosities are included, typically with volumetric ice fractions $<30\%$.

\begin{figure}[htb]
\centerline{\resizebox{3in}{!}{\includegraphics{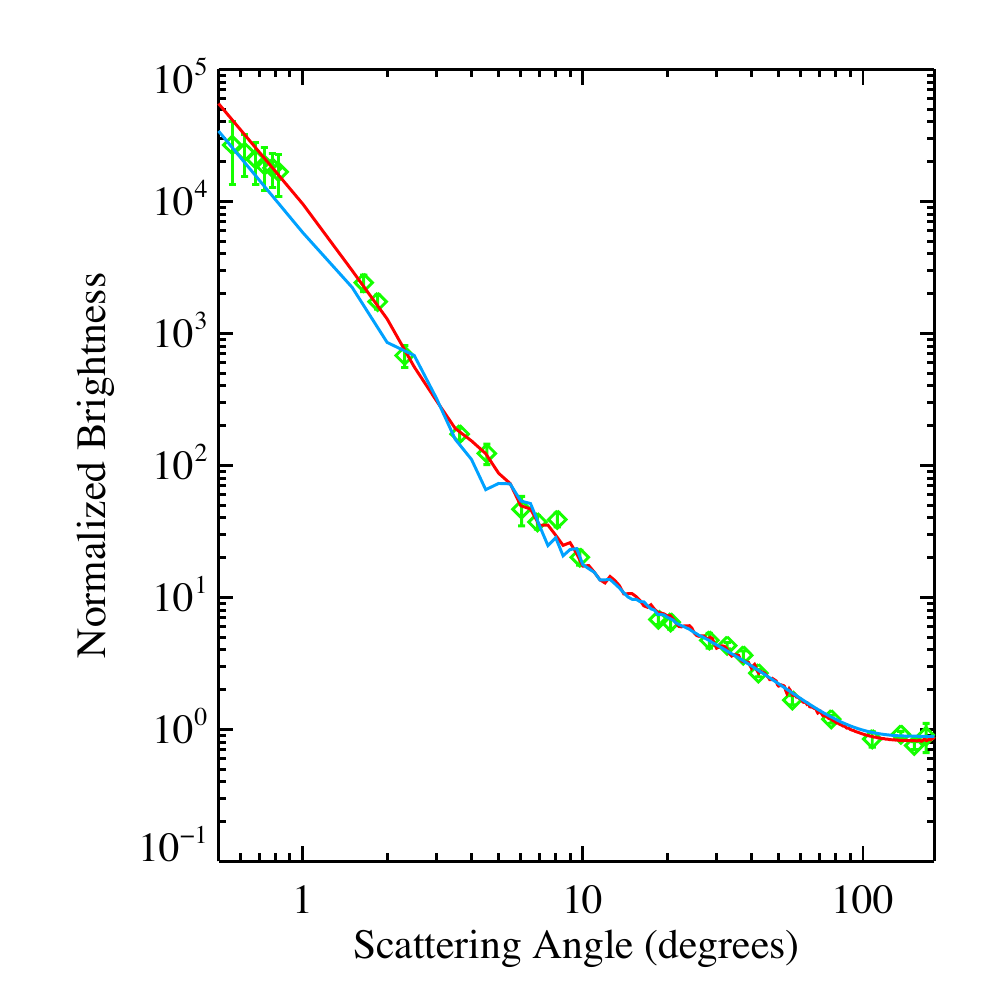}}}
\caption{Two sample Mie theory fits to the D ring SPF.}
\label{mie_theory_fits_D}
\end{figure}

Figure \ref{mie_theory_fits_D} shows the SPFs for two sample fits to the D68 SPF.  The red curve shows the model SPF for the best fit.  The blue curve was drawn from the right 2$\sigma$ island and corresponds to $s_{\rm min} = 35$ and $q = -5.1$.  Both of these fits predict that the forward-scattering peak is 2--3 orders of magnitude brighter at $\theta=0^{\circ}$ than at $\theta \approx 1^{\circ}$. 

\clearpage

\begin{figure}[tb]
\centerline{\resizebox{3in}{!}{\includegraphics{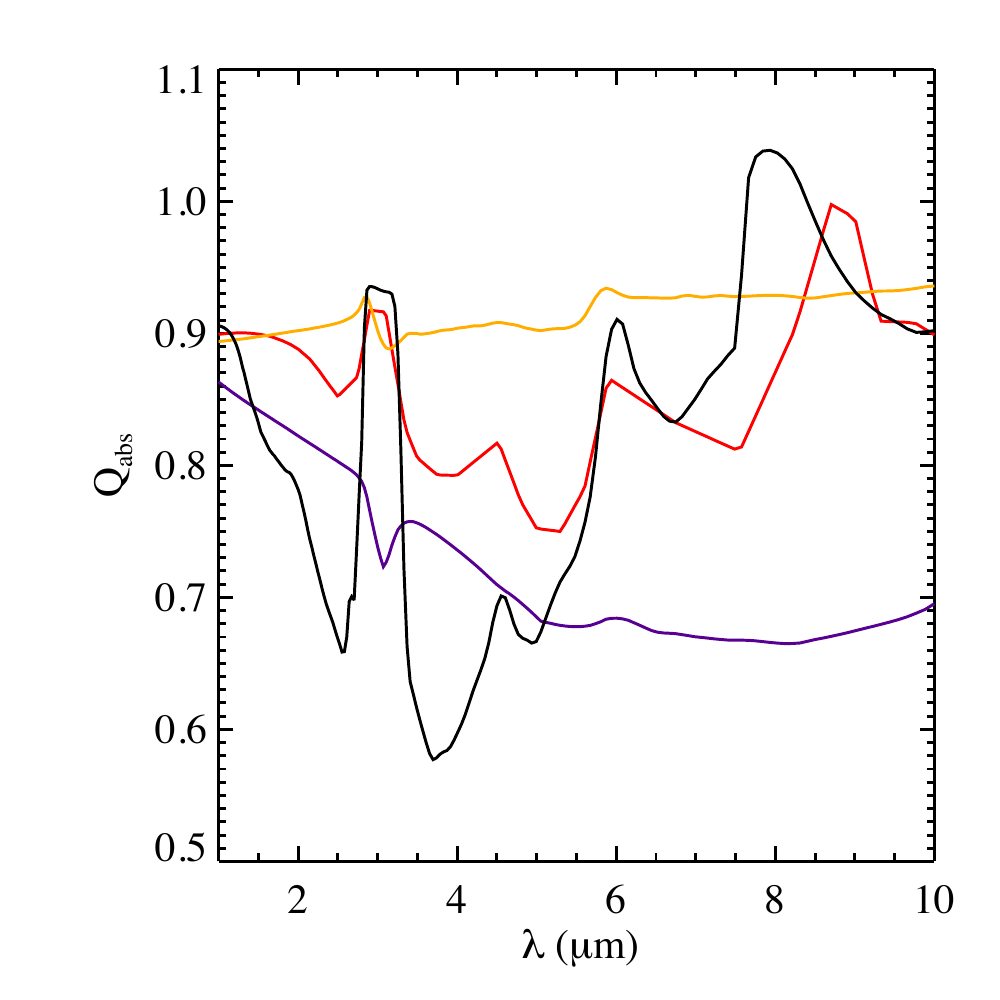}}}
\caption{Predicted absorption coefficients versus wavelength for four of the best-fit
solutions to the G-ring phase function. Red  = 90\% olivine mixed with 10\% water ice, 10\% porosity, 4 - 460 micron grains, $q$ = -2.2 ($\chi^2/\nu$=0.673); purple = 40\% iron mixed with 60\% water ice, 50\% porosity, 36 - 100 micron grains, $q$ = -2.0, ($\chi^2/\nu$=0.942); orange = 50\% organics mixed with 50\% water ice, 0\% porosity, 4 - 460 micron grains, $q$ = -2.2 ($\chi^2/\nu$=0.758); black = 90\% iron-rich orthopyroxene mixed with 10\% water ice, 10\% porosity, 36 - 275 micron grains, $q$ = -3.3 ($\chi^2/\nu$=0.879) }
\label{ir_spec_G}
\end{figure}

Two aspects of these best-fit solutions are surprising and may reflect limitations of Mie-theory-based models: the relatively low fractions of water ice and the relatively high fraction of particles with radii greater than 10 $\mu$m. Let us consider the compositional issues first.  Saturn's dense main rings are known to be composed of relatively pure water ice based on both strong water-ice absorption bands in the near infrared \citep[][and references therein]{cuzzi2009}, as well as the ring's high reflectivity and low emissivity at radio wavelengths \citep[e.g.][]{Pollack1975}. The situation for the dusty rings is less clear, but the F ring and outer D ring both exhibit water-ice absorption bands in the near-infrared \citep{hedman2007, cuzzi2009, hedman2011, vahidinia2011}, and these bands are also visible in spectra of D68 and the G ring (Hedman et al. in prep). Thus we might have expected that the bet-fit solutions would be ones with a higher fraction of water ice than those found above. 
There is evidence for at least two non-icy contaminants in Saturn's rings and moons \citep{cuzzi1998, poulet2003}. One is spectrally neutral and probably corresponds to carbon-rich cometary materials. The other absorbs strongly at wavelength shorter than 0.5 microns and could be either organic compounds or nano-phase iron-rich grains \citep{cuzzi2009, clark2012}. The concentration of these contaminants in D68 and the G ring are currently unknown, and so the lower water-ice fractions in the best-fit models may be due to these contaminants. However, it is also possible that models with substantial non-icy components are favored because these materials suppress structures in the SPF at high scattering angles  that Mie theory predicts for ice-rich spheres (e.g. those responsible for rainbows) that are not seen in the data because the particles have irregular shapes. Studies of the rings' near-infrared spectra should help clarify these issues. For example, Figure~\ref{ir_spec_G} shows the absorption coefficient $Q_{abs}$ versus wavelength for few of the better-fitting G-ring models. The large variations in the $Q_{abs}$ curves should produce obvious features in the rings' near-infrared spectra, and so spectral measurements should reveal whether any of these compositional models are sensible. In the meantime, both these Mie-theory based calculations and the previous comparisons with empirical measurements indicate that the SPFs alone cannot provide strong constraints on the particle composition of these dusty systems.

Turning to the particle-size distributions, we may note that none of the best-fit models have an effective average particle size in the range of a few microns, which would be consistent with the Fraunhofer-diffraction-based analysis. instead, all the best-fit models require a substantial population of particles that are several tens of microns across. These large particles have forward-scattering
lobes that are less than a degree wide according to standard diffraction theory, which is too narrow to be easily detectable in the available data. Thus a substantial population of such large particles would be essentially invisible to the Fraunhofer-based calculations presented above. While Cassini has detected large particles in the G ring \citep{hedman2007}, there are reasons to be skeptical of the idea that most of the particles in these rings are greater than 50 microns across. 

Recall that according to classical Fraunhofer diffraction theory, a particle size distribution with power-law index $q$ should have an SPF that scales like $\theta^{-(q+5)}$. Hence the steepest parts of the forward-scattering lobe, where the brightness goes like $\theta^{-3}$  (see Figure~\ref{dgphase}), are consistent with the shallow size distribution ($q\simeq-2$) favored by the best-fitting Mie-theory models. However, the observed SPFs also become much less steep for scattering angles less than 1$^\circ$ or 2$^\circ$, which is most naturally explained as a cut-off in the size distribution. The best-fitting Mie-theory-based models do not show a sharp reduction in the slope of the SPF at small scattering angles, so these pure power-law models may not be capturing this aspect of the particle size distribution. Furthermore, this difference between the models and the data indicates that Mie theory calculations are not favoring models with many large particles because this improves the fit near the top of the forward-scattering peak. Instead, these models probably favor a substantial population of large grains because such particles can alter the shape of the SPF at larger scattering angles. In particular, large particles will make significant contributions at large scattering angles while making only a small change to the forward-scattering peak between $\theta=0.5^\circ$ and 10$^\circ$. Thus the large particle population probably improves the fit by increasing the flux at large scattering angles relative to the forward-scattering peak. If this is the case, then the amount of large particles could well be  overestimated because Mie theory tends to underestimate the amount of light scattered at moderate to high scattering angles by irregular particles \citep{pollack1980}.


We expect that a more consistent picture of the particle-size distribution will be obtained is we allow for more complex particle size distributions and/or non-spherical particles. Such calculations are beyond the scope of this work, but based on the above considerations, we expect that the final particle size distribution will be rather shallow up to some critical particle size between a few and a few tens of microns.

\section{Implications of the SPFs for exoplanetary debris disks.}
\label{discussion}

Even if we cannot yet  derive a fully consistent physical model for the SPFs of the G ring and D68, our  3-component HG fits still provide an accurate description of these rings' scattering phase functions, and even these simple phenomenological models are useful for evaluating data from exoplanetary disks.
For example, HG functions are commonly used to estimate the degree of forward scattering of a debris disk.  Reported values of $g$ typically range from $0.0$ to $0.3$ for debris disks \citep[e.g.][]{kalas2005,schneider2006,debes2008,thalmann2011}.  However, these estimates are commonly made by fitting the flux ratio along the projected minor axis of the disk and are limited to a small range of scattering angles.  Therefore, they may be a poor representation of the shape of the SPF and true degree of forward scattering.  Recent fits to the shape of the derived SPF suggest significantly more forward scattering \citep{stark2014}.

\begin{figure}[tb]
\centerline{\resizebox{5in}{!}{\includegraphics{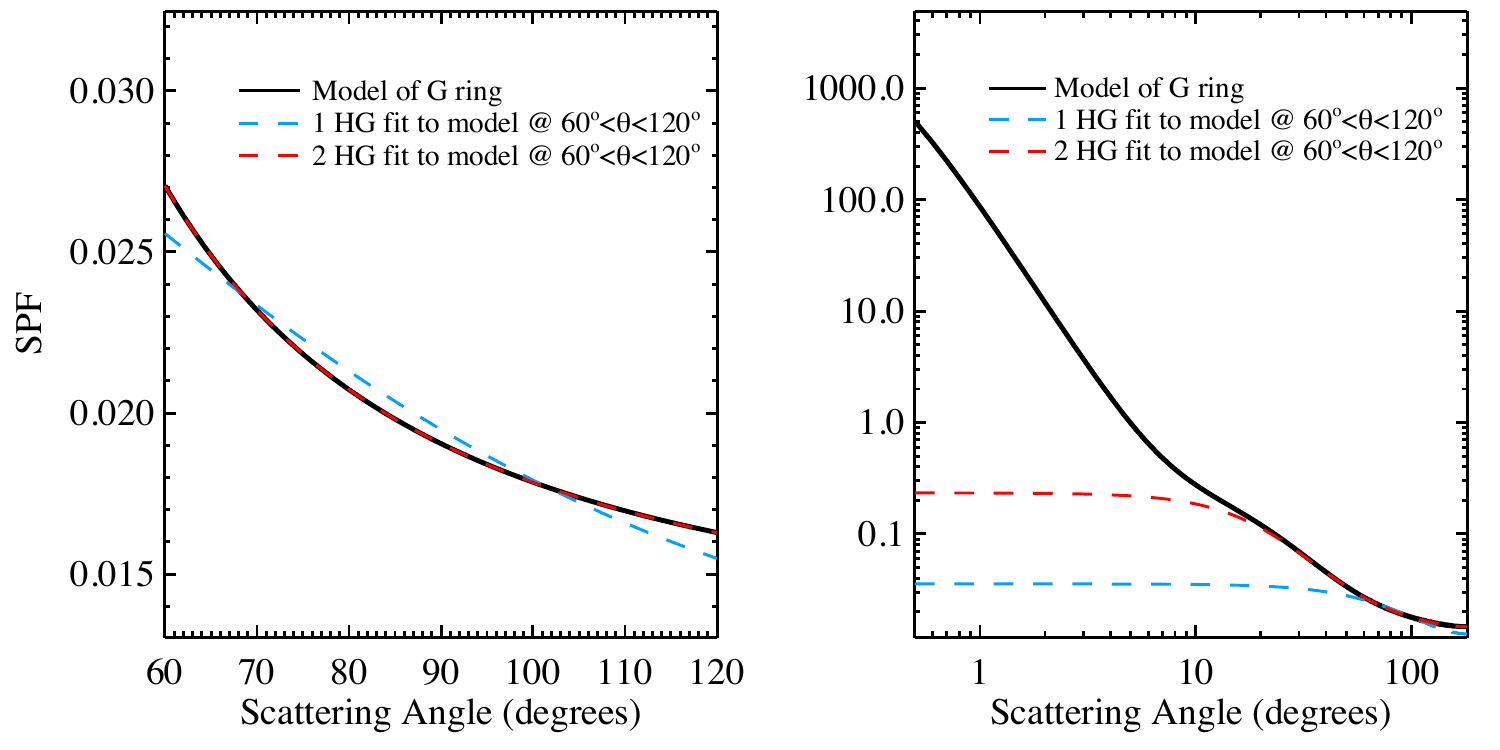}}}
\caption{\emph{Left:} Model G ring SPF (black) over a range of scattering angles observable for a moderately inclined debris disk.  A single HG fit (blue) gives $g=0.17$, on par with estimates of observed debris disks. A 2-component HG fit (red) gives $g_1=0.665$, $w_1=0.505$, $g_2=0.035$, and reproduces the observable SPF well. \emph{Right:} Neither HG fit over the observable range of scattering angles accurately predicts the forward scattering peak at $\theta<10^{\circ}$.}
\label{observable_SPF}
\end{figure}

If a debris disk's true SPF resembled the measured G ring SPF shown in Figure \ref{dgphase}, what might we expect to observe?  A debris disk inclined by 30 degrees from face-on (e.g., HD 181327), would enable observations ranging in scattering angle from 60$^{\circ}$--120$^{\circ}$.  Figure \ref{observable_SPF} shows the variation in the best-fit 3-component HG SPF for the G ring over this range of scattering angles, properly normalized such that the integral of the scattering phase function over all angles is unity.  We adopt this as our G ring SPF ``model."

The left panel of Figure \ref{observable_SPF} shows that when fitting the observable portion of the model SPF with a single HG function, one can roughly reproduce the flux ratio at the smallest and largest scattering angles, but the fit to the shape of the SPF is poor.  The best fit $g=0.17$ is similar to $g$ values reported in the literature for a number of debris disks.  We conclude that over the range of observable scattering angles, the measured SPFs of Saturn's G ring and D68 are roughly consistent with typical debris disk observations.

The left panel of Figure \ref{observable_SPF} also shows the best-fit 2-component HG function, which does very well at reproducing the model over the observable scattering angles.  However, as shown in the right panel, neither HG function fit accurately predicts the model SPF at $\theta<10^{\circ}$.  We conclude that fits to debris disk SPFs, over typical ranges of observable scattering angles, cannot accurately predict the degree of forward scattering. \emph{The reported degrees of forward scattering for debris disks therefore may be greatly underestimated.}

If the model G ring SPF is in fact representative of debris disk SPFs, what would this imply about debris disks?  First, we note that the model G ring SPF near $\theta\approx90^{\circ}$ is $\sim0.02$, when properly normalized such that its integral over all solid angles is unity.  In comparison, isotropic scatterers have a SPF equal to $1/4\pi \sim 0.08$ at all scattering angles.  Thus, debris disks would appear a factor of $\sim4$ dimmer than their isotropically-scattering counterparts, potentially explaining the low apparent albedo of some disks \citep[e.g.,][]{krist2010,golimowski2011,lebreton2012}.

Second, the very forward-scattering model G ring SPF suggests that edge-on disks may be problematic for future exoEarth imaging missions.  \citet{stark2015} showed that cold debris disks may produce a ``pseudo-zodiacal" haze of forward-scattered starlight if oriented within a few tens of degrees of edge-on.  At a projected separation of 1 AU, this ``pseudo-zodi" becomes non-negligible under the assumption of a single HG SPF with $g>0.7$.  The G ring model SPF satisfies this forward-scattering criterion, as the dominant component of the model fit has $g=0.995$.

\section{Conclusions}
\label{conclusions}

\begin{itemize}
\item
The Cassini spacecraft provides measurements of the dusty rings' relative brightness over a large range of scattering angles from 0.5$^\circ$ to  170$^\circ$. These data are consistent with previous studies of dusty rings and debris disks, but also provide new information about the SPF of dusty debris systems at small scattering angles.

\item
The brightness data for these rings is well described by a combination of 3 Henyey-Greenstein functions. If one's goal is to quickly approximate the SPFs of the G ring or D68, we recommend using this fit. 

\item
None of the best fitting compositions from the Amsterdam Light Scattering Database resemble the icy material that we'd expect to find in the G ring or D68.  Also, Mie-thoery-based calculations with a range of compositions fit the phase curves equally well. This suggests that a wide variety of compositions can produce similar SPFs.  Thus, the SPF by itself is probably not a good discriminator of debris disk composition.

\item
Our Fraunhofer fits suggest effective grain radii on the order of a few microns, while the Mie-theory based calculations assuming a power-law size distribution favor typical particles of a few tens of microns.  The applicability of Fraunhofer diffraction theory and Mie Theory to these systems of irregular grains mean that these estimates may be off by factors of a few. 

\item
The strong forward scattering peaks of these dusty systems suggest that the degree of forward scattering in extrasolar debris disks may be greatly underestimated, which could have implications for the albedos of these systems and for future exo-Earth searches.
\end{itemize}

\acknowledgments{We wish acknowledge the Imaging Team, Cassini Project and NASA for the data used for this study. MMH also wishes to acknowledge support from the Cassini Data Analysis and Participating Scientist Program Grant Number NNX14AO27G.}


\end{document}